%% file: main.tex
\documentclass[preprint,12pt]{elsarticle}

\input{includes/packages}
\input{includes/glossary}
\input{includes/todosetup}

\input{includes/macros}

\journal{the Special Issue of EAAI}

\begin{document}

\begin{frontmatter}



\title{Causal Process Mining from Relational Databases\\with Domain Knowledge}


\author[inst1]{Philipp Waibel\corref{cor1}}
\ead{philipp.waibel@wu.ac.at}
\author[inst2]{Lukas Pfahlsberger}
\ead{lukas.pfahlsberger@hu-berlin.de}
\author[inst2]{Kate Revoredo}
\ead{kate.revoredo@hu-berlin.de}
\author[inst2]{Jan Mendling}
\ead{jan.mendling@hu-berlin.de}

\address[inst1]{Institute for Data, Process and Knowledge Management, \\Vienna University of Economics and Business (WU), Vienna, Austria}
\address[inst2]{Humboldt-Universität zu Berlin, Berlin, Germany}

\begin{abstract}
The plethora of algorithms in the research field of process mining builds on directly-follows relations. Even though various improvements have been made in the last decade, there are serious weaknesses of these relationships. Once events associated with different objects that relate with a cardinality of 1:N and N:M to each other, techniques based on directly-follows relations produce spurious relations, self-loops, and back-jumps. This is due to the fact that event sequence as described in classical event logs differs from event causation.
In this paper, we address the research problem of representing the causal structure of process-related event data. To this end, we develop a new approach called {\em Causal Process Mining}. This approach renounces the use of flat event logs and considers relational databases of event data as an input. More specifically, we transform the relational data structures based on the {\em Causal Process Template} into what we call {\em Causal Event Graph}. We evaluate our approach and compare its outputs with techniques based on directly-follows relations in a case study with an European food production company. Our results demonstrate the benefits for enriching process mining with additional knowledge from the domain. 
\end{abstract}



\begin{keyword}
Process mining \sep Causal process mining \sep Causal event graphs \sep Directly-follows graphs \sep Causal knowledge \sep Representational bias
\end{keyword}

\end{frontmatter}


\input{sections/10_introduction}

\input{sections/20_background}

\input{sections/30_approach/30_approach}

\input{sections/40_evaluation/40_evaluation}

\input{sections/50_discussion}

\input{sections/60_conclusion}



\section*{Acknowledgement}
\noindent This work was partly supported by the Einstein Foundation Berlin [grant number EPP-2019-524, 2022] and the WU Wien via the WU-Projects (BAWAG-Stiftung, Projekt-IA 27001663, 2021).
\appendix
\section{Evaluation Results}
\label{sec:evaluation_results}
\noindent Table \ref{tab:pm4py_expected_unexpected_quantity}, \ref{tab:noreja_expected_unexpected_quantity}, and \ref{tab:quantity_withoutviolation_violations} contain the results of the quantitative analysis (cf. \Cref{subsec:metrics-evaluation}) that compares PM4Py with the Noreja Approach.
\input{sections/40_evaluation/expected_unexpected_table_pm4py}
\input{sections/40_evaluation/expected_unexpected_table_noreja}
\input{sections/40_evaluation/noviolation_violation_table_noreja}

\bibliographystyle{template/elsarticle-num-names}

\end{document}

%% file: includes/packages.tex
\usepackage{amsmath}

\usepackage[dvipsnames]{xcolor}
\usepackage{algorithm}
\usepackage{algorithmicx}
\usepackage{algpseudocode}
\usepackage{graphicx}

\usepackage[obeyFinal]{todonotes}

\usepackage{glossaries}
\usepackage{amssymb} 
\usepackage[inline]{enumitem}
\usepackage{tikz}
\usepackage{booktabs}
\usepackage[super]{nth}
\usepackage{hyperref}
\usepackage{cleveref}
\usepackage{listings}
\usepackage{pifont}
\usepackage{soul}
\usepackage{svg}
\usepackage{comment}
\usepackage{outlines}
\usepackage{tcolorbox}
\usepackage{xspace}
\usepackage{rotating}
\usepackage{adjustbox}

\usepackage{multirow}
\usepackage{lscape}

\usepackage{caption}
\usepackage{subcaption}

\definecolor{LightGray}{rgb}{0.97,0.97,0.97}

\crefname{lstlisting}{listing}{listings}
\Crefname{lstlisting}{Listing}{Listings}

\crefname{definition}{definition}{definition}
\Crefname{definition}{Definition}{Definition}

\usepackage{siunitx}

%% file: includes/glossary.tex
\newacronym{cpm}{CPM}{Causal Process Mining}

\newacronym{pes}{PES}{Prime Event Structure}
\newacronym{ceg}{CEG}{Causal Event Graph}
\newacronym{aceg}{ACEG}{Aggregated Causal Event Graph}
\newacronym{dfg}{DFG}{Directly-Follows Graph}
\newacronym{pm}{PM}{Process Mining}
\newacronym{bpm}{BPM}{Business Process Management}
\newacronym{cpt}{CPT}{Causal Process Template}

\newacronym{kpi}{KPI}{Key Performance Indicator}
\newacronym{xoc}{XOC}{eXtensible Object-Centric}

\newacronym{arti}{ARTI}{Artifact Type Level Interaction}
\newacronym{ari}{ARI}{Artifact Instance Level Interaction}
\newacronym{ddg}{DDG}{Dynamic Dependence Graph}

\newacronym{er}{ER}{Entity Relationship}
\newacronym{pk}{PK}{Primary-Key}
\newacronym{fk}{FK}{Foreign-Key}

\newacronym{it}{IT}{Information Technology}
\newacronym{xml}{XML}{eXtensible Markup Language}

\newacronym{bpms}{BPMS}{Business Process Management System}
\newacronym{bpmn}{BPMN}{Business Process Model and Notation}
\newacronym{xor}{XOR}{eXclusive OR}
\newacronym{xes}{XES}{eXtensible Event Stream}
\newacronym{epc}{EPC}{Event-driven Process Chain}
\newacronym{ui}{UI}{user interface}

%% file: includes/todosetup.tex
\definecolor{darkCyan}{RGB}{0,145,207}
\definecolor{lightCyan}{RGB}{73,186,235}

\definecolor{darkBlue}{RGB}{69,92,221}
\definecolor{lightBlue}{RGB}{108,129,253}

\definecolor{darkViolet}{RGB}{118,68,211}
\definecolor{lightViolet}{RGB}{149,109,224}

\definecolor{darkRed}{RGB}{195,37,37}
\definecolor{lightRed}{RGB}{229,95,69}

\definecolor{darkOrange}{RGB}{233,105,20}
\definecolor{lightOrange}{RGB}{255,143,43}

\definecolor{darkBrown}{RGB}{167,88,35}
\definecolor{lightBrown}{RGB}{198,124,69}

\definecolor{darkGreen}{RGB}{92,147,21}
\definecolor{lightGreen}{RGB}{115,192,17}

\definecolor{darkTeal}{RGB}{76,131,114}
\definecolor{lightTeal}{RGB}{114,173,155}

\definecolor{darkYellow}{RGB}{255,225,6}
\definecolor{lightYellow}{RGB}{255,246,128}

\definecolor{darkGrey}{RGB}{141,156,163}
\definecolor{lightGrey}{RGB}{178,194,203}

\definecolor{lightTeal}{RGB}{50,193,186}



%% file: includes/macros.tex
\newenvironment{iiilist}%
{\begin{enumerate*}[label=(\roman*)]}%
{\end{enumerate*}}

\specialcomment{scribble}{\color{blue}}{\color{black}}

\newtheorem{definition}{Definition}

\newcommand{\pre}[1]{{\bullet\,{#1}}}
\newcommand{\post}[1]{{#1}\bullet}

\newcommand{\STAB}[1]{\begin{tabular}{@{}c@{}}#1\end{tabular}}

%% file: sections/10_introduction.tex
\section{Introduction}
\label{sec:introduction}
\noindent 
\gls{pm} is a research field focusing on the development of automatic analysis techniques that provide insights into business processes based on event data \citep{DBLP:books/sp/Aalst16}. Over the last decade, \gls{pm} research has spent considerable efforts on improving algorithms for automatic process discovery. Notable advancements of recent algorithms include the inductive miner~\cite{leemans2018scalable}, the evolutionary tree miner~\cite{buijs2014quality}, or the split miner~\cite{augusto2019split}. These algorithms differ in their balance between fitness, simplicity, generalization, and precision in order to better cope with different event log characteristics or use cases \citep{DBLP:books/sp/Aalst16,augusto2018automated}, but they have in common that they build on event logs as input and directly-follows relations for generating models.

Even though most \gls{pm} techniques build on event logs and directly-follows relations, they exhibit some serious weaknesses. Once events associated with different objects that relate with a cardinality of 1:N and N:M to each other, techniques based on directly-follows relations produce spurious relations, self-loops, and back-jumps. The incapability to handle such cardinality constitutes a representational bias \citep{Aalst2011OnTR, van2011process} that causes problems for appropriately representing, a.o., logistic processes~\citep{gerke2009case}. Here, interacting objects cannot be traced using a single case identifier. Recent works on artifact-related \gls{pm} \citep{DBLP:journals/tsc/LuNWF15} or object-centric \gls{pm} \citep{DBLP:conf/sefm/Aalst19, DBLP:conf/caise/LiMCA18} as well as approaches that consider relational databases \citep{DBLP:journals/kais/MurillasRA20, 10.1007/978-3-030-46633-6_2, ANDREWS2020113265} or multi-dimensional event data \citep{DBLP:journals/corr/abs-2005-14552} address the problem by associating a single event with one or more objects. However, they do not exploit causal domain knowledge such as visible in the foreign key relationships of the databases storing event data. This is a fundamental problem, because it is not possible to reconstruct causal relationships from observation of behavior alone~\cite{pearl2019seven}. As a consequence, techniques based on \glspl{dfg} produce connections that are not causal, yielding overly complex models.

In this paper, we address the problem of representing the causal structure of process-related event data. To this end, we develop a new approach called {\em \gls{cpm}}. This approach renounces the use of flat event logs and considers relational databases of event data as well as causal knowledge as an input. More specifically, we transform the relational data structures based on the {\em \gls{cpt}} into what we call {\em \gls{ceg}}. In turn, these \glspl{ceg} can be stored in a graph database and they can be aggregated to models that we call \emph{\glspl{aceg}}. \gls{cpm} exploits these structures and offers multiple operations to investigate the process from various perspectives and at different levels of aggregation. With the {\em \gls{cpt}}, we propose a novel component that helps to integrate alternative knowledge sources. 
Moreover, we evaluate \gls{cpm} and compare its outputs with techniques based on directly-follows relations in a case study with an European food production company. Our results demonstrate that directly-follows miners produce a large number of spurious relationships, which our approach captures correctly. 

The rest of the paper is structured as follows. Section~\ref{sec:background} illustrates the fundamental problem of \glspl{dfg} by making use of a production scenario example. 
Based on this discussion, we derive requirements for representing the corresponding class of scenarios appropriately. 
Section~\ref{sec:approach} defines our approach of \gls{cpm}. 
We illustrate how our approach handles cardinality, causality, and process instance aggregation. 
Section~\ref{sec:evaluation} conducts an evaluation by comparing the \gls{cpm} approach to common \gls{dfg} approaches as well as discusses the requirements in the context of related work. Section~\ref{sec:conclusion} concludes the paper with a summary and an outlook on future research.

%% file: sections/20_background.tex
\section{Background}
\label{sec:background}

\noindent
In this section, we illustrate the research problem and relate it to Pearl's three-level causal hierarchy \cite{pearl2019seven}. Upon this basis, we identify four requirements for addressing the problem of representing the causal structure based on event data. Using these requirements, we compare contributions from prior research.

\subsection{Problem statement}
\label{subsec:problem-statement}
\noindent This section describes the problem of \gls{pm} approaches that use event logs with a single case identifier for generating \glspl{dfg}. Extracting directly-follows relations and representing them as a graph is the ``de facto standard''~\citep[p. 324]{van2019practitioner} of both commercial tools and the main share of the academic implementations. 

Let us look at the different technical steps of conducting a \gls{pm} analysis according to this de facto approach. 
Event logs stem from transactions that are executed by humans or by software systems. 
These transactions are typically persisted in tables of relational databases, which themselves are connected by foreign-key relations of different cardinalities. 
As a preparation for \gls{pm}, a process analyst creates an event log by defining the obligatory case identifier, specifying the process activities, and related properties such as resources, timestamps, or business objects~\citep{jans2019building}. The result is a flattened event log in the form of a single file, persisted table, or non-persisted view. Taking this event log as an input, a DFG-based algorithm generates a process model, e.g. using the heuristic miner~\citep{DBLP:books/sp/Aalst16} or any other miner that uses directly-follows relations alone or in combination with more complex ones~\citep{DBLP:journals/is/CiccioMMM18}. These algorithms are efficient in constructing directly-follows graphs, but entail fundamental interpretation problems that are ``considered harmful''~\citep[p. 323]{van2019practitioner}. We illustrate these problems using an example of one of our industry partners. 

\begin{figure*}[t]
	\centering
	\includegraphics[width=0.9\columnwidth]{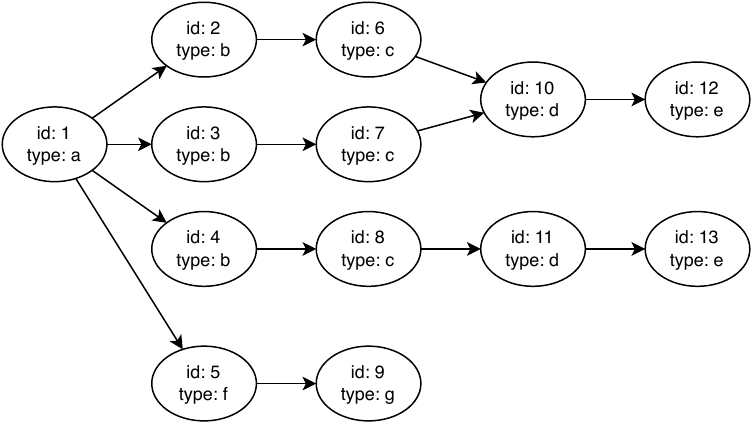}
	\caption{Example of an order received by our partner company.}
	\label{fig:exampleProblem}
\end{figure*}

Our partner company is producing and delivering food to major supermarket chains. Their order-to-cash process is triggered by supermarket orders ($a$). These orders are broken down into a list of item suborders ($b$). Each of these items are separately picked from the warehouse ($c$), packaged up ($d$) and sent in one or multiple deliveries ($e$). Each order also triggers the emission of an invoice ($f$), which is settled by a payment ($g$). Figure \ref{fig:exampleProblem} depicts an example of an order.

This description of the order-to-cash process exhibits the causal structure in terms of which event triggers which subsequent event. Let us assume for the moment that we do not have any prior knowledge about causalities. Given is an event log of the process containing the following sequence recording only a fraction of the event types: $\langle a,c,f \rangle$. 
The observation of this single sequence does not allow us to conclude about the causal structure between the event types except that succeeding events cannot cause an event that had already happened earlier. That means, the evidence equally supports ($a$) causing ($c$) and ($c$) causing ($f$) as much as ($a$) causing ($c$) and ($a$) causing ($f$) or ($a$) and ($c$) combined causing ($f$). This observation does not change if we observe the same sequence ten times, one thousand times or more often. Indeed, in case of our industry partner, the invoicing is triggered by the completion of picking. However, due to typical delays of processing, the invoice is only emitted on the next business day when the delivery to the customer is already done. It should be noted that miners that use directly-follows relations are assuming causality from this sequence of execution.

The problem worsens if we consider cases with multiple instantiation. Problems of case identification and focus shifts have been described earlier~\citep{gerke2009case,gerke2009process}. These problems stem from bundling and unbundling operations. Consider that picking is done in two steps: If an order event ($a$) leads to the creation of three order items ($b$) and their picking from the warehouse ($c$), we might observe an event sequence like $\langle a,b,c,b,b,c,c \rangle$. Clearly, ($b$) does neither cause itself nor is it caused by ($c$), but it results from the initial order with several items. Miners that use directly-follows relations cannot distinguish this situation where ($a$) causes several ($b$) events which each trigger a~($c$) from processes where a failures of ($b$) and ($c$) might cause repetitions.

The consequence of the described discovery problem is that directly-follows-based miners produce relationships that are not causal. For the case with $\langle a,c,f \rangle$, we obtain a relationship representing the wrong causality; for $\langle a,b,c,b,b,c,c \rangle$ even several wrong ones. In fact, only a subset of the directly-follows relationships are causal: Those from ($a$) to ($b$) and from ($b$) to ($c$). Notably, \gls{pm} techniques based on directly-follows relations
produce relationships that are spurious for the described process (from ($b$) to ($b$), from ($c$) to ($b$), and from ($c$) to ($c$)). Partially, these relationships are responsible for the commonly known \textit{Spaghetti model} problem. Additionally, what is inextricably linked to the problem of spurious relationships is the misleading calculation of performance information related to time and cost. For instance, this leads to incorrect throughput times between two activities~\citep{van2019practitioner}. The question is how we can prevent these spurious relationships from being included in the output of the process miner?

The answer to this question is dissatisfactory for directly-follows miners and given by Pearl in the context of statistical inference~\citep[p. 59]{pearl2019seven}: ``Associations alone cannot identify the mechanisms responsible for the changes that occurred, the reason being that surface changes in observed associations do not uniquely identify the underlying mechanism responsible for the change.'' This observation also holds for \gls{pm}: If miner $X$ constructs model $x$ and miner $Y$ model $y$ both with 100\% fitness, the event data offers no evidence to refute any of these models even if the one with higher precision might look more plausible. Connected to this, another problem occurs which is often denounced by analysts who use these models for process improvement and root-cause analysis: Having models with a high fitness helps to get a realistic picture of the process in a holistic manner. Nevertheless, common miners require input that flattens the process instance relation and maps it to a single case identifier, which in turn inflates the directly-follows graph as an output. 
What is missing are techniques that avoid 
spurious relationships, misleading connections, or inaccurate proportions right from the start. 

Pearl describes a way out of this dilemma by the help of his three-level causal hierarchy~\citep{pearl2019seven}. The levels are distinguished by the queries that they are able to answer. The \nth{1} level refers to associations and queries like ``what is associated with each other?''. Queries at this level can be fully answered based on observations. The \nth{2} level refers to interventions and queries like ``what will happen if we intervene?''. Already at this level it cannot be fully answered based on observational data because it requires understanding of correlations. Level 3 refers to counterfactuals and queries like ``what would have happened if we had not intervened?''. Queries at this level require knowledge of the functional mechanisms that produce observational data. Pearl describes that the limitations of the \nth{1} level can be overcome by combining data with assumptions encoded in a causal model~\citep{pearl2019seven}. In this paper, we will transfer this idea formulated for statistical inference to our approach of \gls{cpm}.

\subsection{Requirements}\label{subsec:Requirements}
\noindent
To address the aforementioned challenges, we identify the following requirements. 

\paragraph{RQ1: Input Data representing Causal Event Structure}
The requirement is concerned with the input data for a \gls{pm} algorithm. Instead of working with event logs defined for a single case notion, we observe the requirement to keep and represent the
causality between events as stored in the source database.
State-of-the-art techniques assume that this causal structure is unknown and typical try to reconstruct causality based on the directly-follows relations observed in the event log~\cite{DBLP:books/sp/Aalst16}.
If the input for a \gls{pm} technique already captures causal structures, this will avoid generating process models that include connection that are actually not causal.

\paragraph{RQ2: External Knowledge about Causal Event Structure}
Based on Pearl's comments, we observe the requirement to integrate domain knowledge about the causality between events into \gls{pm} techniques. This is specifically needed to capture 
events and their causal relationships that can not be observed solely from the event data. 
Integrating this domain knowledge about causality facilitates the comparison with input data. In this way, input data can be validated and violations that contradict the causal structure can be identified. 

\paragraph{RQ3: Event Cardinalities}
We also observe the requirement to explicitly capture the cardinality between events and event types.
The analysis of the cardinality is a key requirement for those processes that exhibit bundling and unbundling scenarios~\citep{gerke2009case,gerke2009process}, processes with divergence and convergence~\citep{DBLP:conf/sefm/Aalst19}, or batching operations~\cite{DBLP:conf/icpm/WaibelNBRM20}. 
The lack of support by classic \gls{pm} techniques leads to the construction of loops and spurious relationships for event types that are in a 1:N relationship with the event type that triggers a process.

\paragraph{RQ4: Aggregation of Causal Event Structures}
We observe that \gls{pm} techniques are required for constructing aggregated models from input based on causal event structures. A corresponding \gls{pm} technique has to construct representations that support the analysis at a coarse-granular level of a process model as well as at the detailed level of a process instance. To this end, different business objects have to be selected as a focus of analysis. 
This would allow an analyst to inspect a business process at different levels of abstraction~\citep{DBLP:conf/sac/WeberFMS15}.

\subsection{Prior Research}\label{subsec:Prior_Research}
\noindent
Next, we provide an overview of related work. We specifically focus on the extent to which existing approaches address or implement the four requirements we identified.
We classify prior research that addressed the problem of causal relations and cardinality between events when learning a process model in two streams of research: 
\begin{iiilist}
\item process discovery from relational databases where the causal relation and the cardinality are provided by the data structure; and
\item approaches that infer the casual relation and cardinality information from flat event logs and then discover the process model.
\end{iiilist}

For what concerns stream {\itshape i)}, \citet{DBLP:conf/caise/LiMCA18} create an object-centric event log format, named \gls{xoc}, that does not require a case notion as required for the XES format. \citet{DBLP:conf/caise/LiMCA18} argue that this object-oriented event log format helps to store relationships in the form of 1:N and N:M as it is common in databases. The problems of classic flattened event logs are also discussed in~\citep{DBLP:conf/sefm/Aalst19} in which an object-centric \gls{pm} approach is presented. A projection of this event log considering the different existing objects allows for the discovery of the corresponding process model. In \cite{DBLP:journals/fuin/AalstB20}, an approach for discovering object-centric petri nets is presented. Moreover, \citet{8456350} suggest a method for identifying correlation between events considering the path between objects. 

Pursuing a slightly different direction, \citet{DBLP:conf/apn/Fahland19} presents formal semantics for processes with N:M interactions considering unbounded dynamic synchronization of transitions, cardinality constraints, and history-based correlation of token identities. In~\citep{DBLP:conf/bpm/EsserF19} and \citep{DBLP:journals/corr/abs-2005-14552}, the cardinality is captured by the concept of one event being part of multiple cases by using \emph{labeled property graphs}. 
\citet{DBLP:journals/corr/abs-2005-14552,DBLP:conf/bpm/EsserF19} transform event logs into graphs to store structural and temporal relationships between events. They discuss how edges between events define a causal relationship, based on the assumption that events are related to each other if there is an underlying entity to which both events belong. They later extend \emph{labeled property graphs} to \emph{event knowledge graphs} by integrating ``further knowledge'' like entity inference, resources, or actors \citep[p. 285ff]{fahland2022process}.

\citet{10.1007/978-3-030-46633-6_2} propose an approach for learning models that represent multiple viewpoints of an event log stored in a relational database. These models are annotated with frequency and performance measures supporting the analysis of bottlenecks without the notion of a case. From these models, it is possible to derive event logs and reconstruct process models using standard \gls{pm} tools. 
In \cite{DBLP:journals/corr/abs-2103-07184}, process cube operations such as slice and dice are applied to object-centric event logs for deriving object-centric sublogs. Furthermore, \citet{DBLP:journals/tsc/LuNWF15} generate a process model for each artifact out of event data extracted from a relational database. The foreign-key relationships are used to grasp causal dependencies between the events and define the order of the events in the event log. However, when discovery techniques are applied to the standard event log generated, the causal relation is not explicit and therefore not considered for the process model generation. A relation between the process models associated to the artifacts is discovered based on the cardinalities observed in the relational database. In this way, they are able to identify Artifact Type Level Interaction and Artifact Instance Level Interaction.

Table \ref{tb_RQRelatedWork_Stream_i} summarizes how each of these prior works of stream {\itshape i)} meets the four requirements we identified above. ($\bullet$) indicates partial support. 

\begin{table}[h]
\caption{Specification of how the current approaches address the requirements of stream~{\itshape i)}}
\label{tb_RQRelatedWork_Stream_i}
\begin{tabular}{p{7,0cm}p{1,1cm}p{1,1cm}p{1,1cm}p{1,1cm}} 
\toprule
\textbf{Approach} & \textbf{RQ1} & \textbf{RQ2} & \textbf{RQ3} & \textbf{RQ4} \\\midrule
\citet{DBLP:journals/fuin/AalstB20}         &    $\bullet$& & $\bullet$     &    \\
\citet{DBLP:conf/sefm/Aalst19}       &  $\bullet$&   &  $\bullet$   &      \\
\citet{10.1007/978-3-030-46633-6_2}      &   $\bullet$&  &  $\bullet$   &    \\
\citet{DBLP:journals/tsc/LuNWF15}       &  $\bullet$ &  &  $\bullet$   &     \\
\citet{DBLP:journals/corr/abs-2103-07184}      & $\bullet$ &   &     & $\bullet$    \\
\citet{DBLP:conf/caise/LiMCA18}     &   $\bullet$&  &  $\bullet$   &     \\
\citet{8456350}   &   $\bullet$ & ($\bullet$) &   $\bullet$  &     \\
\citet{DBLP:conf/apn/Fahland19}   &   $\bullet$ & &   $\bullet$  &     \\
\citet{DBLP:conf/bpm/EsserF19}&   $\bullet$&  &  $\bullet$   &  $\bullet$   \\
\citet{DBLP:journals/corr/abs-2005-14552}  &   $\bullet$ & &  $\bullet$   & $\bullet$    \\
\citet{fahland2022process}  &   $\bullet$ & ($\bullet$) &  $\bullet$   & $\bullet$    \\
\bottomrule
\end{tabular}
\end{table}

For what concerns stream {\itshape ii)}, \citet{7095577} identify dependencies between activities based on control or data flow for discovery. A dynamic dependence graph captures all activities of the event log as well as their dependencies. The process model is then derived from this graph. In \cite{DBLP:journals/eswa/DiamantiniGPA16}, techniques for identifying causality in the event log and filtering are used to learn instance graphs. \citet{DBLP:conf/otm/LuFASWHA17} identify causality between events by modeling the traces as a partial order of its events \cite{leemans2022partial}. The approach then uncovers patterns describing directly-cause, eventually-cause, and concurrence relations between events. The concurrence relation is defined by pairs of events that are not eventually-causing one another. \citet{DBLP:conf/apn/DumasG15} discuss a \gls{pm} approach based on \glspl{pes}. Their approach transforms the cases in an event log into \glspl{pes} and then use the concept of asymmetric event structures to create a process model.

Ponce de Le{\'{o}}n et al. use a similar approach~\cite{10.1007/978-3-319-24953-7_4} that creates labeled partial orders from the event log and independence information about the activities, merge them in an event structure, which is then transformed into an occurrence net. A process model is derived from folding the occurrence net. Furthermore, \citet{DBLP:conf/icpm/Bergenthum19} also uses \glspl{pes} as an intermediate step to create a process model. The approach learns concurrency between events and use it for deriving a partial order of the event log to yield its \gls{pes}. Approaches by \citet{DBLP:conf/sac/WeberFMS15} and by \citet{DBLP:journals/is/ConfortiDGR16} present ideas that identify and hide subordinate processing paths and integrate them into multiple instance activities in the discovered models. 

\begin{table}[h]
\caption{Specification of how the current approaches address the requirements of stream~{\itshape ii)}}
\label{tb_RQRelatedWork_Stream_ii}
\begin{tabular}{p{7,0cm}p{1,1cm}p{1,1cm}p{1,1cm}p{1,1cm}} 
\toprule
\textbf{Approach} & \textbf{RQ1} & \textbf{RQ2} & \textbf{RQ3} & \textbf{RQ4} \\\midrule
\citet{7095577}  & $\bullet$& & & \\
\citet{DBLP:journals/eswa/DiamantiniGPA16} & $\bullet$ &  & &\\
\citet{DBLP:conf/apn/DumasG15} & $\bullet$& & & $\bullet$\\
\citet{DBLP:conf/otm/LuFASWHA17} & $\bullet$& & & \\
\citet{10.1007/978-3-319-24953-7_4}  & $\bullet$& & &$\bullet$ \\
\citet{DBLP:conf/icpm/Bergenthum19}  & $\bullet$& & &$\bullet$ \\
\citet{DBLP:conf/sac/WeberFMS15}     & $\bullet$& & $\bullet$ & $\bullet$ \\
\citet{DBLP:journals/is/ConfortiDGR16}& & & & $\bullet$\\ 
\bottomrule
\end{tabular}
\end{table}

Table \ref{tb_RQRelatedWork_Stream_ii} summarizes how each of these prior work of stream {\itshape ii)} meets the requirements we defined in Section \ref{subsec:Requirements}.

Almost all these works have in common that they make a best effort for reconstructing the causal relationships between the events and activities of the underlying multi-perspective business process. Nevertheless, as Pearl emphasizes, the level of \textit{intervention} and \textit{counterfactual} queries is not reachable for those purely data-driven approaches since they only contain ``purely observational information'' \citep[p. 56]{pearl2019seven}. We address this problem in the next section by developing a \gls{cpm} approach that we call \textit{Noreja}\footnote{Noreja is the Celtic goddess of mining. It is associated with the East alps region, where the Celtic kingdom of Noricum existed before Roman conquest. The author team was jointly established in Vienna at the heart of this region when starting to work on this research.}, which is based on Causal Event Graphs (\glspl{ceg}) and Aggregated Causal Event Graphs (\glspl{aceg}). We use a \gls{cpt} to internalize external knowledge based on causal assumptions of domain experts as as well as meta information from relational databases (e.g., cardinalities like 1:N and N:M)~\citep{pearl2019seven}. 

%% file: sections/30_approach/30_approach.tex
\section{An Approach to Causal Process Mining}
\label{sec:approach}
\noindent 
In this section, we introduce our \gls{cpm} approach and explain how it differentiates itself from classical approaches. Then, we summarize preliminaries from prior research, provide definitions of the key formal notions, define formal operations, and describe visualizations.

\subsection{Setting the Broader Context}
\begin{figure*}[t]
	\centering
	\includegraphics[width=1.0\columnwidth]{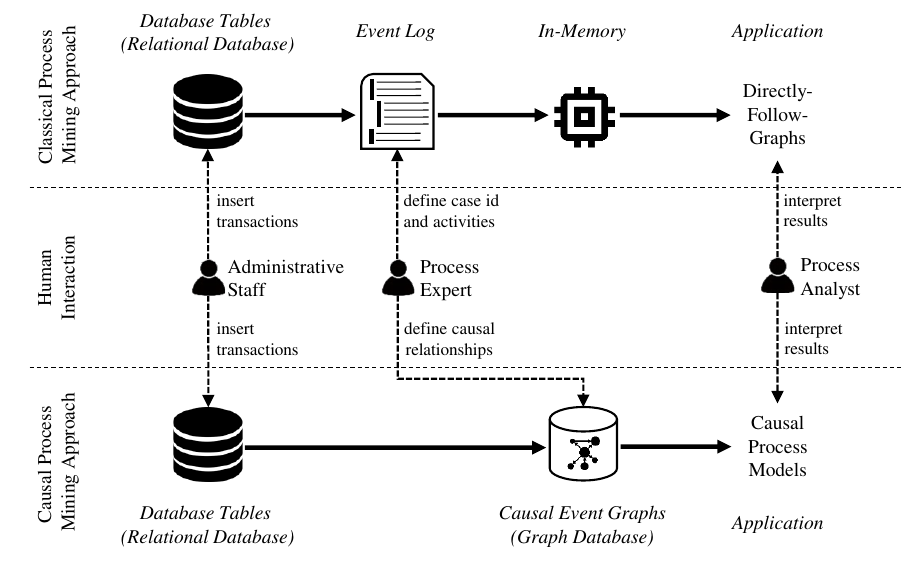}
	\caption{Differences between \emph{causal} and \emph{classical} (event log based) \gls{pm}}
	\label{fig:causalapproachdesc}
\end{figure*}

\noindent Before going into the details of our approach, we want to juxtapose it with classical \gls{pm} approaches based on \glspl{dfg} and event logs.
Figure \ref{fig:causalapproachdesc} gives an overview of the different steps of each approach and where humans are involved. Classical approaches (top lane in Figure~\ref{fig:causalapproachdesc}) build on four steps. First, transactions are inserted into a database during the execution of a business process. Those transactions are persisted inside tables of relational databases, which themselves are connected by foreign-key relations with different cardinalities. Second, in preparation for creating the event log, a process expert defines the obligatory case identifier, specifies the process activities and related properties, such as resources, timestamps, or other business objects. Based on this information, a flattened event log is extracted. Third, the event log is loaded into memory where the \gls{dfg} is calculated as an output. Fourth, a process analyst can interpret the visualization on the screen.

Our \gls{cpm} approach (bottom lane in Figure \ref{fig:causalapproachdesc}) extracts the relevant data from a relational database, which matches the classical approach. However, the second step has fundamental differences, because the extraction of a dedicated event log is skipped. Instead, we directly transfer the relational database structures into \glspl{ceg} and store them in a graph database. In this way, we avoid flattening the data, such that we keep cardinalities and causal relationships between data objects. The user-facing application then only has to visualize the \glspl{ceg} in a use-case specific manner. From a technical perspective, this shifts computation from the application layer to the database layer. We obtain reusability as different user-facing applications can query the data independently, and flexibility 
as the \gls{ceg} is not bound to a predefined case identifier. In the subsequent sections, we present our approach from formal perspective.

\input{sections/30_approach/33_graph_construction}
\input{sections/30_approach/34_graph_analysis}

%% file: sections/30_approach/33_graph_construction.tex
\subsection{Noreja Approach}
\label{subsec:graph-construction}
\noindent 
The Noreja Approach for \acrfull{cpm} consists of the following consecutive steps:
\begin{iiilist}
    \item First, a set of relevant tables 
    is selected from the source relational database and joined over foreign keys.
    \item Then, the \acrfull{cpt} is defined. 
    \item Afterwards, each causally connected tuple in the joined tables is transformed to a so-called \acrfull{ceg} by merging fragments that share common events. 
    \item Based on the \gls{ceg}, different aggregations, called \gls{aceg}, are calculated for each primary key. 
    \item Eventually, the \glspl{ceg} and the \glspl{aceg} are used to analyze the processes.
\end{iiilist}
Next, we discuss each step in turn. 

\subsubsection{Input Data Selection}
\label{subsec:input-data-selection}
\noindent
The first step of the Noreja Approach is concerned with determining which tables of a relational database are required for analysis. To this end, we recall the definition of a database schema based on Calvanese et al~\cite{calvanese2019formal}.

\begin{definition}[Relational Schema]
Let $R$ be the set of relations of a relation database. A relational schema of a relation $r \in R$ is a pair $\langle r.name,r.attrs \rangle$, where $r.name$ is the relation name and $r.attrs$ is a nonempty tuple of attributes. The first attribute in $r.attrs$ must be the identifier denoting the primary key ($r.pk$) followed by identifiers of other relations denoting foreign keys ($r.FK=\{r.fk_1,...,r.fk_n\}$), when existing, a time ($r.timestamp$) registering a time event, and lastly other attributes. We call $|r.attrs|$ the arity of $r$.
\end{definition}

\begin{definition}[Catalog]
 Let $R$ be the set of relations of a relation database. A catalog $Cat$ is any subset of relations ($Cat \subseteq R$) where every two relations $r_i,r_j \in Cat$ have different primary keys ($r.pk_i \neq r.pk_j$) in their relational schemas.
\end{definition}


A catalog of relations $Cat$ may be selected by a process analyst and its relations joined using the \emph{left outer join} $\bowtie_L$.

\begin{definition}[Left Outer Join, Causally Connected Tuple]
Let $Cat= \{ r_1,\dots,r_n \}$ be a catalog. Left outer join is a relational schema $r_J = r_1 \bowtie_L r_2 \dots r_{n-1} \bowtie_L r_n$ such that for each $r_i \in Cat$, there exists another $r_j \in Cat$ with $j\neq i$, such that either $\exists r_j.fk \in r_j.FK$ where $r_j.fk = r_i.pk$ or $\exists r_i.fk \in r_i.FK$ where $r_i.fk = r_j.pk$. We call a row in $r_J$ a causally connected tuple.
\end{definition}

\begin{figure}[t]
    \centering
    \includegraphics[width=1.0\columnwidth]{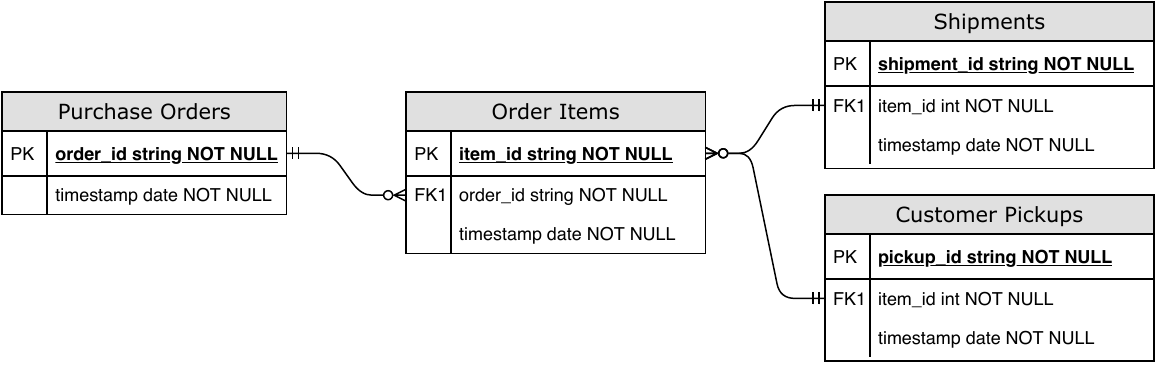}
    \caption{Example \acrshort{er}-diagram. \emph{Note:} To simplify the diagram, we have omitted additional properties, e.g., customer name, item type, shipment address, etc.}
    \label{fig:ceg_construction_er}
\end{figure}

As an example, \Cref{fig:ceg_construction_er} depicts an \acrshort{er}-diagram of a catalog from an order-to-cash process, $Cat=\{$ $Purchase \, Orders,$ $Order \, Items,$ $Shipments,$ $Customer \, Pickups\}$.
The \emph{Purchase Orders} table holds the order received events, i.e., an event is persisted in the table as soon as an order is received. 
The order is then broken down into a list of items, which are then separately picked from the warehouse. 
These warehouse picking events are stored in \emph{Order Items}. 
As can be seen in the ER-Diagram, each of these \emph{Order Items} is linked to a \emph{Purchase Order} via a foreign-key relationship. 
Eventually, the items are shipped to the customer, these events are stored in the \emph{Shipments} table, or directly picked up by the customer, these events are stored in the \emph{Customer Pickups} table. 
Again these events are related to the corresponding \emph{Order Items} via foreign-key relationships.
We can define the left outer join $r_J$ over these tables using the foreign-key relationships. 
A \emph{causally connected tuple} contains, for instance, one \emph{order\_id} indicating the purchase order, one \emph{item\_id} indicating a corresponding order item, and either the corresponding \emph{shipment\_id} or \emph{pickup\_id}. Mind that if identifiers in connected tables do not exist, the left outer join will include $null$ values.

\subsubsection{Causal Process Template Definition}
\noindent
The left outer join over the selected tables has yielded a relation, in which each row is a causally connected tuple. Our approach requires a process analyst to specify a partial order over the selected relations. We call such a  partial order \acrfull{cpt}.
\begin{definition}[Causal Process Template]
Let $Cat= \{ r_1,\dots,r_n \}$ be a catalog. A \acrfull{cpt} is a partial order $\prec~\subseteq Cat \times Cat$ over the relations of the catalog.
\end{definition}

The \gls{cpt} is the basis for constructing the causal relationships over the activities of a process. It can be defined based on the \acrshort{er}-diagram or on an adaptation of it. In our example, a customized \gls{cpt} could define that the \emph{Customer Pickups} is a causal successor of \emph{Purchase Orders} and not of \emph{Order Items} as it is defined in the \acrshort{er}-Diagram.
After the catalog and the \gls{cpt} are defined, the \acrfull{ceg} is constructed.

\input{sections/30_approach/331_ceg_transformation}
\input{sections/30_approach/332_ceg_aggregation}

%% file: sections/30_approach/331_ceg_transformation.tex
\subsubsection{\acrlong{ceg} Construction}
\label{subsec:graph-construction}
\noindent
A \acrfull{ceg} provides the means to model events and the causal relationship between them. An event and a \gls{ceg} are defined as follows.
\begin{definition} [Event] Let $Cat= \{ r_1,\dots,r_n \}$ be a catalog. An event $e_i^v$ of a relation $r_i \in Cat$ is a tuple $e_i^v=<r_i.pk^v,r_i.timestamp^v>$ which indicates the value of the attributes $r_i.pk$ and $r_i.timestamp$, i.e., which is the identifier of the event and when it happened. An event $e_i^v$ is associated with a tuple of the table $r_i$.
\end{definition}

\begin{definition} [\acrlong{ceg}]  Let $Cat= \{ r_1,\dots,r_n \}$ be a catalog, $\prec \subseteq Cat \times Cat$ the \gls{cpt} over $Cat$, $r_J = r_1 \bowtie_L r_2 \dots r_{n-1} \bowtie_L r_n$ the left outer join of the relations in $Cat$ considering the defined \gls{cpt}, and $rj$ a causally connected tuple of $r_J$. A \acrlong{ceg} is a tuple  $CEG = \langle E, \leq, \lambda \rangle$, where $E \subseteq \{e_1^{{rj}_1},\dots,e_1^{{rj}_n},\dots , e_n^{{rj}_1},\dots,e_n^{{rj}_n}\}$, $\leq~\subseteq E \times E$ is a binary relation where $(e_i^{rj},e_j^{rj}) \in \leq$ if $r_i \prec r_j$, and $\lambda : E \to ET$ is the labeling function that defines the event type of an event.   
\end{definition}
For each causally connected tuple $rj$ of $r_J$, a \gls{ceg} ($CEG_{rj}$) may be created where $E$ is the set of events associated with the tuple $rj$. 
The set $r_J$ of all causally connected tuples $rj$ of a relational database catalog $Cat$ induces the Causal Event Database $CEG_{DB} =  \bigcup_{rj \in r_J} CEG_{rj}$. For that a $CEG_{rj}$ is called a \emph{fragment} of $CEG_{DB}$. Two or more fragments may have common events, for instance, when two orders share one delivery. In this case, the fragments are connected and the shared event is called a \emph{batching event}~\cite{DBLP:conf/icpm/WaibelNBRM20}. Overall, $CEG_{DB}$ is typically disconnected and the maximally connected components define a partition of $CEG_{DB}$. Accordingly, we define two functions:
\begin{itemize}
    \item $Comp(CEG_{DB})$: it returns the set of all components of $CEG_{DB}$, and
    \item $Frag(CEG_{DB})$: it returns the set of all fragments $CEG_{rj}$ of $CEG_{DB}$ where $rj \in r_J$. 
\end{itemize}

As an example, \Cref{fig:ceg_construction_database} shows an instantiation of the relational database defined in \Cref{fig:ceg_construction_er}. According to the Noreja Approach, we construct the following \glspl{ceg} fragments (\Cref{fig:ceg_construction_nodes} depicts them graphically). For simplification, we used the identifier value as the label of the events. 
\begin{itemize}
    \item $CEG_1 = \langle E_1, \leq_1, \lambda_1 \rangle$ with
    \begin{itemize}
        \item $E_1 = \{or1, pi1, pi2, pi3, sh1, sh2\}$, 
        \item $\leq_1 = \{(or1, pi1)$, $(or1, pi2)$, $(or1, pi3)$, $(pi1, sh1)$, $(pi2, sh1)$, \\$(pi3, sh2)\}$, and
        \item $\lambda_1 = \{(or1,Receive\; Purchase\; Order)$, $(pi1,Pick\; Order\; Item)$,\\ $(pi2,Pick\; Order\; Item)$, $(pi3,Pick\; Order\; Item)$,\\ $(sh1,Register\; Shipment)$, $(sh2,Register\; Shipment)\}$       
    \end{itemize}
    \item $CEG_2 = \langle E_2, \leq_2, \lambda_2 \rangle$ with 
    \begin{itemize}
        \item $E_2 = \{or2, pi4, sh2\}$, 
        \item $\leq_2 = \{(or2, pi4)$, $(pi4, sh2)\}$, and
        \item $\lambda_2 = \{(or2,Receive\; Purchase\; Order)$, $(pi4,Pick\; Order\; Item)$,\\ $(sh2,Register\; Shipment)\}$
    \end{itemize}
    \item $CEG_3 = \langle E_3, \leq_3, \lambda_3 \rangle$ with
        \begin{itemize}
        \item $E_3 = \{or3, pi5, cu1\}$, 
        \item $\leq_3 = \{(or3, pi5)$, $(pi5, cu1)\}$, and 
        \item $\lambda_3 = \{(or3,Receive\; Purchase\; Order)$, $(pi5,Pick\; Order\; Item)$, \newline $(cu1,Register\; Customer\; Pickup)\}$
    \end{itemize}
\end{itemize}
\begin{figure}[t]
    \centering
    \includegraphics[width=1.0\columnwidth]{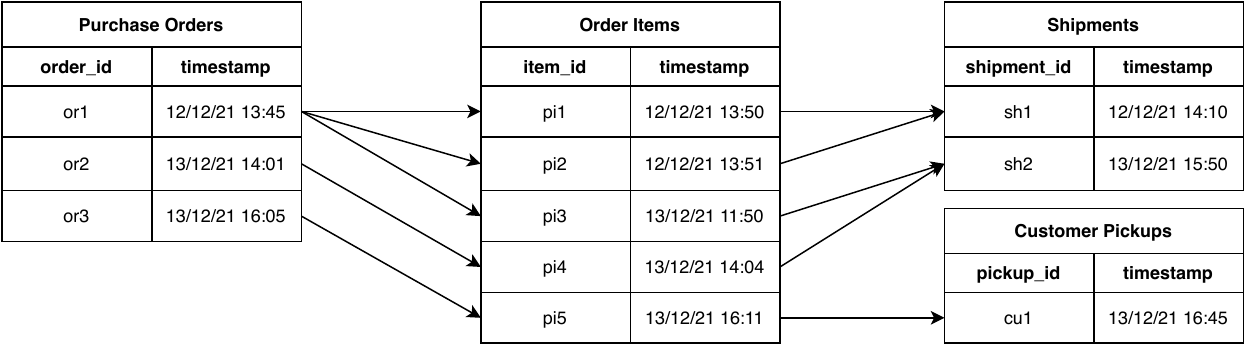}
    \caption{Example instantiation of the database. The arrows depict the relationships between the entries.}
    \label{fig:ceg_construction_database}
\end{figure}
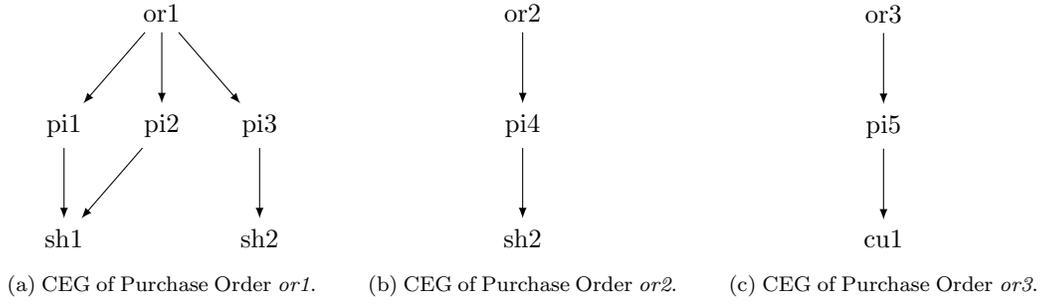
\begin{figure}[t]
    \centering
    \begin{subfigure}[t]{0.3\textwidth}
        \centering
        \tikzset{>=latex}
        \begin{tikzpicture}[node distance=2cm, sibling distance=1.3cm, edge from parent/.style={draw,-latex}, font=\footnotesize]
            \node (or1) {or1}
            child {node (pi1) {pi1} 
                child {node (sh1) {sh1}}
            }
            child {node (pi2) {pi2} }
            child {node (pi3) {pi3} 
                child {node (sh2) {sh2}}
            };
            \draw[->] (pi2) -> (sh1);          
        \end{tikzpicture}
        \caption{\gls{ceg} of Purchase Order \emph{or1}.}
        \label{fig:ceg_construction_nodes_or1}
    \end{subfigure}
    \hfill
    \begin{subfigure}[t]{0.3\textwidth}
        \centering
        \tikzset{>=latex}
        \begin{tikzpicture}[node distance=2cm, sibling distance=2cm, edge from parent/.style={draw,-latex}, font=\footnotesize]
            \node (or2) {or2}
            child {node (pi4) {pi4} 
                child {node (sh2) {sh2}}
            };
        \end{tikzpicture}
        \caption{\gls{ceg} of Purchase Order \emph{or2}.}
        \label{fig:ceg_construction_nodes_or2}
    \end{subfigure}
    \hfill
    \begin{subfigure}[t]{0.3\textwidth}
        \centering
        \tikzset{>=latex}
        \begin{tikzpicture}[node distance=2cm, sibling distance=2cm, edge from parent/.style={draw,-latex}, font=\footnotesize]
            \node {or3}
            child {node (pi5) {pi5} 
                child {node (cu1) {cu1}}
            };
        \end{tikzpicture}
        \caption{\gls{ceg} of Purchase Order \emph{or3}.}
        \label{fig:ceg_construction_nodes_or3}
    \end{subfigure}
    \caption{\glspl{ceg} fragments for the data shown in \Cref{fig:ceg_construction_database}.}
    \label{fig:ceg_construction_nodes}
\end{figure}

Furthermore, we define the following accompanying functions and concepts.
\begin{definition}[Preset and Postset Nodes] 
Let $E$ be a set of events, and $R\subseteq E \times E$ a binary relation over $E$. We define the preset function $\pre e : E \rightarrow \mathcal{P}(E)$ as $\pre{e} = \{ x \in E~|~(x,e)\in R \}$ for an event $e \in E$. As well as the postset function $\post e: E \rightarrow \mathcal{P}(E)$ as $\post{e} = \{x \in E~|~(e,x)\in R\}$ for an event $e \in E$.
\end{definition}
\begin{definition}[Event Type, 
Identifier and Time] 
\label{def:event_type}
Let $E$ be a set of events, and $ET$ the set of event types. We define $\tau : E \to ET$ as the function that returns the event type of an event $e \in E$. We define $e^\tau=\tau(e)$. 
We define $\mathrm{id} : E \to I$ as an index function and $\mathrm{t}: E \to \mathbb{R}$ as a time function. 
\end{definition}

The result of this construction are \glspl{ceg} that are partially disconnected. 
However, fragments are partially connected via batching events e.g., the shipment event \emph{sh2} is shared between the order received event \emph{or1} and \emph{or2}.

To find all batching events of a given \gls{ceg} for analysis purposes, we define the following function.
\begin{definition}[Batching Event] 
\label{def:get_batching_events}
Let $E$ be a set of events. We define a function $\mathrm{batch} : E \to \mathbb{N}$ that returns $1$ when an event is a batch event and $0$ otherwise.
\end{definition}
We also define the following function that returns the cycle time between a given event and its causal predecessors.
\begin{definition}[Event Cycle Time]
\label{def:event_duration}
Let $E$ be a set of events. For each event $e \in E$, we define its cycle time as a function $cyc : E \to \mathbb{N}$ such that \\
\resizebox{.98\hsize}{!}{$cyc(e)=
\begin{cases}
t(e) - \max(\{t(x)| x \in \pre{e} \wedge t(x)\leq t(e)\}), & \text{if } \pre{e} \neq \emptyset \wedge \exists x \in \pre(e) | t(x) \leq t(e)\\
0, &\text{otherwise}
\end{cases}
$}
\end{definition}
As defined in \Cref{def:event_duration}, the cycle time of an event $e$ is the timespan between the event and the latest precedent event, i.e., $\max(t(\pre{e}))$. Mind that this calculation includes both waiting time and processing time.
Also, if the event $e$ is a start event, i.e., $\pre{e} = \emptyset$, there is no precedent event. 
In such situations, different customized options are possible e.g., a 0 or a \emph{NULL} value could be returned. 
Another option is to use estimations like those proposed in~\cite{martin2020retrieving}.

%% file: sections/30_approach/332_ceg_aggregation.tex
\subsubsection{\acrlong{ceg} Aggregation}
\label{subsec:graph-aggregation}
\noindent
In the next step of the Noreja Approach, different aggregation levels of the transformed \glspl{ceg} are created.
Before we discuss the different aggregation levels, we have to define how an aggregation of a \gls{ceg} is created.

As a first step, a set of event types $ET$ is created from all types $e^\tau$ of all events $e \in E$ of the \gls{ceg}.
For instance, when we consider $CEG_1$ from \Cref{subsec:graph-construction} (depicted in \Cref{fig:ceg_construction_nodes_or1}) and that the labels $\lambda_1$ define the event types, then the final set of aggregated event types $ET$ contains \emph{Receive Purchase Order}, \emph{Pick Order Item}, and \emph{Register Shipment}. 

For the aggregation of the \gls{ceg}, we have to calculate the quantity for each event type with the help of the function $c(E,et)$.
\begin{definition}[Event Type Quantity] 
\label{def:event_type_count}
Let $E$ be a set of events, $ET$ a set of event types, and $et \in ET$ an event type. We define the event type count $c:\wp(E) \times ET \to \mathbb{N}$ such that $c(X,et)=|\{ e\in X~|~e^{\tau}=et \}|$, where $X \subseteq \wp(E)$.
\end{definition}
For instance, for $CEG_1$ the resulting quantities for aggregated 
event types $et \in ET$ are:
\begin{itemize}
    \item $c(E_1,Receive\; Purchase\; Order)=1$,
    \item $c(E_1,Pick\; Order\; Item)=3$, and
    \item $c(E_1,Register\; Shipment)=2$.
\end{itemize}  

As a next step, we add a causal relationship between two event types, $et_a, et_b \in ET$, \emph{iff} there exist two events, $e_a, e_b \in E$, in the \gls{ceg} such that $et_a = \tau(e_a)$, $et_b = \tau(e_b)$, and $(e_a, e_b) \in \leq$ in the \gls{ceg}. 
For instance, for $CEG_1$ this would be the case for the event types \emph{Receive Purchase Order} and \emph{Pick Order Item}, and \emph{Pick Order Item} and \emph{Register Shipment}.

Also for the aggregated relationships the quantities are calculated. 
This is done via the function $cr(R,et_1,et_2)$ defined in \Cref{def:event_type_rel_count}.
\begin{definition}[Event Type Relationship Quantity] 
\label{def:event_type_rel_count}
Let $E$ be a set of events, $R\subseteq E \times E$ a binary relation over $E$, $ET$ a set of event types, and $et_1, et_2 \in ET$ event types. We define the quantity of the relationship between event types $cr: R \times ET \times ET \to \mathbb{N}$ such that 
$cr(R,et_1,et_2)=|\{ v \in E~|~(v, w)\in R,v^t=et_1, w^t=et_2 \}|$
\end{definition}
For $CEG_1$ (cf. \Cref{fig:ceg_construction_nodes_or1}) the following relationship quantities are calculated: \begin{itemize}
    \item $cr(\leq,Receive\; Purchase\; Order, Pick\; Order\; Item)=1$,
    \item $cr(\leq,Pick\; Order\; Item, Register\; Shipment)=3$.
\end{itemize}

As a last step, the outgoing and incoming cardinalities of each relationship are calculated. 
For this, we first define the functions $\mathrm{Out}(v,et,R)$ (\Cref{def:relationship_out_cardinality}) and $\mathrm{In}(v,et,R)$ (\Cref{def:relationship_in_cardinality}). 
\begin{definition}[Event to Event Type Outgoing Cardinality] 
\label{def:relationship_out_cardinality}
Let $E$ be a set of events, $R\subseteq E \times E$ a binary relation over $E$, and $ET$ a set of event types. We define 
the number of outgoing relationships of an event $v \in E$ to an event type $et \in ET$ as $\mathrm{Out}:E \times ET \times \wp(R) \to \mathbb{N}$ such that $\mathrm{Out}(v,et,R)=|\{(v, w)\in R~|~w^t=et \}|$.
\end{definition}
This means that the function $\mathrm{Out}(v,et,R)$ calculates the output degree of an event $e \in E$ to a particular event type $et \in ET$. For instance, for $CEG_1$ the resulting output degrees are:
\begin{itemize}
    \item $\mathrm{Out}(or1, \text{\emph{Pick Order Item}}, \leq_1) = 3$ and 
    \item $\mathrm{Out}(pi3, Register\; Shipment, \leq_1) = 1$.
\end{itemize}

\begin{definition}[Event to Event Type Incoming Cardinality] 
\label{def:relationship_in_cardinality}
Let $E$ be a set of events, $R\subseteq E \times E$ a binary relation over $E$, and $ET$ a set of event types. We define 
the number of incoming relationships to an event type $et \in ET$ from an event $v \in E$ as $\mathrm{In}:ET \times E \times \wp(R) \to \mathbb{N}$ such that $\mathrm{In}(et,v,R)=|\{ (v, w)\in R~|w^t=et \}|$.
\end{definition}

By this definition, the function $\mathrm{In}(et, v,R)$ calculates the input degree from an event type $e^t$ to an event $e \in E$. In $CEG_1$ the resulting cardinalities are:
\begin{itemize}
    \item $\mathrm{In}(\text{\emph{Pick Order Item}}, or1, \leq_1) = 1$ and 
    \item $\mathrm{In}(Register\; Shipment, pi3, \leq_1) = 1$.
\end{itemize}

To be able to describe the cardinality between two event types, we have to extend the functions from \Cref{def:relationship_out_cardinality} and \Cref{def:relationship_in_cardinality} with the possibility to get the minimum and maximum of $\mathrm{Out}(v,et,R)$ and $\mathrm{In}(et, v,R)$. 
To this end, we first define the functions to calculate the minimum and maximum for the outgoing cardinality, \Cref{def:relationship_min_max_out_cardinality}, and then the functions to calculate the same for the incoming cardinality in \Cref{def:relationship_min_max_in_cardinality}.
\begin{definition}[Minimum and Maximum Outgoing Cardinality] 
\label{def:relationship_min_max_out_cardinality}
Let $E$ be a set of events, $R\subseteq E \times E$ a binary relation over $E$, $ET$ a set of event types, and $et_1,et_2 \in ET$ event types. We define  
\begin{itemize}
    \item the minimum outgoing cardinality from one event type to another event type as $\mathrm{MinOut}:\wp(R) \times ET \times ET \to \mathbb{N}$ such that $\mathrm{MinOut}(R,et_1,et_2)=\min(\{ out(v,et_2,R)~|~v \in E \wedge  v^t=et_1 \})$, and 
    \item the maximum outgoing cardinality from one event type to another event type as $\mathrm{MaxOut}:\wp(R) \times ET \times ET \to \mathbb{N}$ such that $\mathrm{MaxOut}(R, et_1,et_2)=\max(\{ out(v, et_2, R)~|~v \in E \wedge v^t=et_1 \})$. 
\end{itemize}
\end{definition}
When we apply these functions $MinOut$ and $MaxOut$ to $CEG_1$, we get the following results for the relationship: 
\begin{itemize}
    \item between \emph{Receive Purchase Order} and \emph{Pick Order Item}\newline
    $MinOut(\leq_1, Receive\; Purchase\; Order,Pick\; Order$ $Item)=3$ and \newline 
    $MaxOut(\leq_1,Receive\; Purchase\; Order,Pick\; Order\; Item)=3$ , and 
    \item between \emph{Pick Order Item} and \emph{Register Shipment}\newline
    $MinOut(\leq_1, Pick\; Order\; Item, Register\; Shipment)=1$ and \newline 
    $MaxOut(\leq_1, Pick\; Order\; Item, Register\; Shipment)=1$ .
\end{itemize}

\begin{definition}[Minimum and Maximum Incoming Cardinality] 
\label{def:relationship_min_max_in_cardinality}
Let $E$ be a set of events, $R\subseteq E \times E$ a binary relation over $E$, $ET$ a set of event types, and $et_1,et_2 \in ET$ event types. We define  
\begin{itemize}
    \item the minimum incoming cardinality from one event type to another \\event type as $\mathrm{MinIn}:\wp(R) \times ET \times ET \to \mathbb{N}$ such that $\mathrm{MinIn}(R, et_1,et_2)=\min(\{ in(et_1, v, R)~|~ v \in E \wedge v^t=et_2 \})$, and 
    \item the maximum incoming cardinality from one event type to another event type as $\mathrm{MaxIn}:\wp(R) \times ET \times ET \to \mathbb{N}$ such that $\mathrm{MaxIn}(R, et_1,et_2)=\max(\{ in(et_1, v, R)~|~v \in E \wedge v^t=et_2 \})$. 
\end{itemize}
\end{definition}
By apply the functions $MinIn$ and $MaxIn$ to $CEG_1$ we get the following resulting cardinalities: 
\begin{itemize}
    \item $MinIn(\leq_1, Pick\; Order$ $Item, Receive\; Purchase\; Order)=1$ and 
    \item $MaxIn(\leq_1, Pick\; Order\; Item, Receive\; Purchase\; Order)=1$, and 
    \item $MinIn(\leq_1, Register\; Shipment, Pick\; Order\; Item)=1$ and 
    \item $MaxIn(\leq_1, Register\; Shipment,Pick\; Order\; Item)=2$.
\end{itemize}
Summarizing, we define an aggregation of a \gls{ceg}, called a \gls{aceg}, the following way:
\begin{definition} [\acrlong{aceg}] 
\label{def:aceg}
A tuple $ACEG = \langle ET, \leq, \circ, \bowtie, \rtimes, \ltimes, \lambda \rangle$ is a labeled \emph{\acrfull{aceg}}, 
where $ET$ is a set of event types, 
$\leq$ is the causality binary relation on $ET$, 
$\circ : ET \to \mathbb{N}$ defines the quantity of the events of a event type, 
$\bowtie : \leq \to \mathbb{N}$ defines the quantity of the causality relations, 
$\rtimes : \leq \to \mathbb{N} \times \mathbb{N}$ defines the minimum and maximum incoming cardinality of the causality relations, 
$\ltimes : \leq \to \mathbb{N} \times \mathbb{N}$ defines the minimum and maximum outgoing cardinality of the causality relations, and 
$\lambda : ET \to \Lambda$ is the labeling function, 
such that $\leq$ is a partial order. 
\end{definition}
As an example, the formal specification of the \gls{aceg} of $ceg_1$ is defined by $ACEG_1 = \langle ET, \leq, \circ, \bowtie, \rtimes, \ltimes, \lambda \rangle$ with 
\begin{itemize}
    \item $ET = \{or, pi, sh\}$, 
    \item $\leq = \{(or, pi)$, $(pi, sh)\}$, 
    \item $\circ = \{(or, 1)$, $(pi, 3)$, $(sh, 2)\}$, 
    \item $\bowtie = \{(or, pi, 1)$, $(pi, sh, 3)\}$,  
    \item $\rtimes = \{(or, pi, 1, 1)$, $(pi, sh, 1, 2)\}$,
    \item $\ltimes = \{(or, pi, 3, 3)$, $(pi, sh, 1, 1)\}$, and
    \item $\lambda = \{(or,Receive\; Purchase\; Order)$, $(pi,Pick\; Order\; Item)$,\\ $(sh,Register\; Shipment)\}$
\end{itemize}

\begin{figure}[t]
    \centering
    \tikzset{>=latex}
    \begin{tikzpicture}[level distance=2.8cm, text width=3cm, align=center, edge from parent/.style={draw,-latex}, font=\footnotesize]
        \node (or1) {Receive \\Purchase Order\\(1)}
        child {
            node (pi1) {Pick Order Items\\(3)} 
            child {
                node (sh1) {Register \\Shipment\\(2)}
                edge from parent node[left=-0.6, font=\footnotesize] {1..2 : 1..1\\(3)}
            }
            edge from parent node[left=-0.6, font=\footnotesize] {1..1 : 3..3 \\ (1)}
        };
    \end{tikzpicture}
    \caption{Aggregation of of $CEG_1$, from \Cref{subsec:graph-construction}. Node and relationship quantity in parenthesis and relationship cardinality in the form \{$\rtimes_{min}$..$\rtimes_{max}$ : $\ltimes_{min}$..$\ltimes_{max}$\}. }
    \label{fig:aceg_construction}
\end{figure}
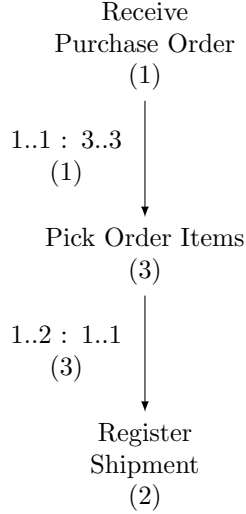

\Cref{fig:aceg_construction} depicts the resulting \gls{aceg} graphically. 
In the figure, the node and relationship quantities, i.e., $\circ$ and $\bowtie$, are shown in parenthesis. For simplification, the cardinality functions, i.e., $\rtimes$ and $\ltimes$, are represented in the format \{$\rtimes_{min}$..$\rtimes_{max}$ : $\ltimes_{min}$..$\ltimes_{max}$\} in which the first pair of elements, i.e., $\rtimes_{min}$..$\rtimes_{max}$, represent the minimum and maximum of the incoming cardinality and the second pair, i.e., $\ltimes_{min}$..$\ltimes_{max}$, the minimum and maximum of the outgoing cardinality.

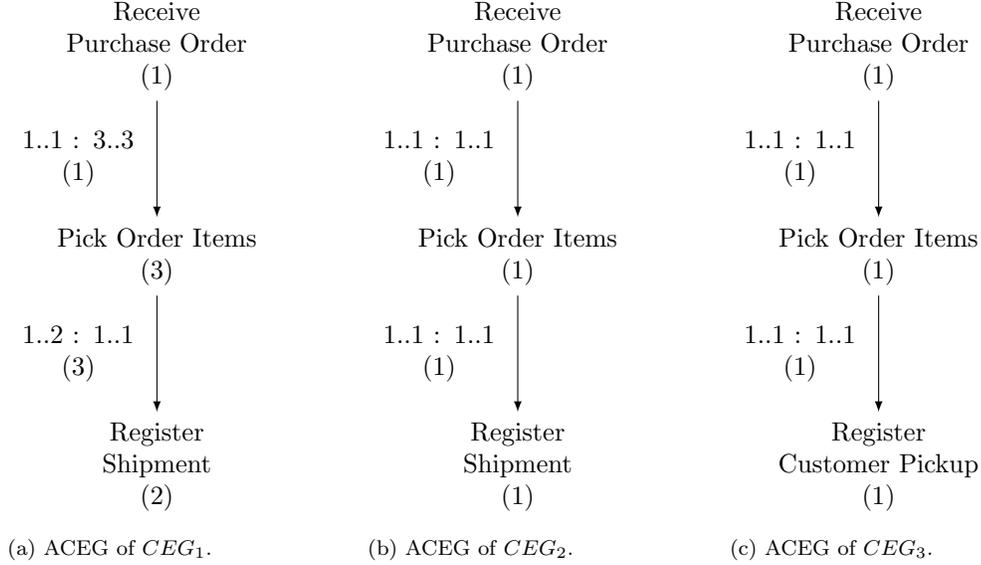
\begin{figure}[t]
    \centering
    \begin{subfigure}[t]{0.3\textwidth}
        \centering
        \tikzset{>=latex}
        \begin{tikzpicture}[level distance=2.8cm, text width=3cm, align=center, edge from parent/.style={draw,-latex}, font=\footnotesize]
            \node (or1) {Receive \\Purchase Order\\(1)}
            child {
                node (pi1) {Pick Order Items\\(3)} 
                child {
                    node (sh1) {Register \\Shipment\\(2)}
                    edge from parent node[left=-0.6, font=\footnotesize] {1..2 : 1..1\\(3)}
                }
                edge from parent node[left=-0.6, font=\footnotesize] {1..1 : 3..3 \\ (1)}
            };
        \end{tikzpicture}
        \caption{\gls{aceg} of $CEG_1$.}
        \label{fig:aggregation_level_1_1}
    \end{subfigure}
    \hfill
    \begin{subfigure}[t]{0.3\textwidth}
        \centering
        \tikzset{>=latex}
        \begin{tikzpicture}[level distance=2.8cm, text width=3cm, align=center, edge from parent/.style={draw,-latex}, font=\footnotesize]
            \node (or2) {Receive \\Purchase Order\\(1)}
            child {
                node (pi4) {Pick Order Items\\(1)} 
                child {
                    node (sh2) {Register \\Shipment\\(1)}
                    edge from parent node[left=-0.6, font=\footnotesize] {1..1 : 1..1\\(1)}
                }
                edge from parent node[left=-0.6, font=\footnotesize] {1..1 : 1..1\\(1)}
            };
        \end{tikzpicture}
        \caption{\gls{aceg} of $CEG_2$.}
        \label{fig:aggregation_level_1_2}
    \end{subfigure}
    \hfill
    \begin{subfigure}[t]{0.3\textwidth}
        \centering
        \tikzset{>=latex}
        \begin{tikzpicture}[level distance=2.8cm, text width=3cm, align=center, edge from parent/.style={draw,-latex}, font=\footnotesize]
            \node (or2) {Receive \\Purchase Order\\(1)}
            child {
                node (pi4) {Pick Order Items\\(1)} 
                child {
                    node (co1) {Register \\Customer Pickup\\(1)}
                    edge from parent node[left=-0.6, font=\footnotesize] {1..1 : 1..1\\(1)}
                }
                edge from parent node[left=-0.6, font=\footnotesize] {1..1 : 1..1\\(1)}
            };
        \end{tikzpicture}
        \caption{\gls{aceg} of $CEG_3$.}
        \label{fig:aggregation_level_1_3}
    \end{subfigure}
    \caption{\nth{1} aggregation level of the \gls{ceg} depicted in \Cref{fig:ceg_construction_nodes}.}
    \label{fig:aggregation_level_1}
\end{figure}

The Noreja Approach provides three different levels of aggregation, which allows the analysis of the process on three different levels of detail:

\paragraph{Aggregation Level 1}
The \nth{1} aggregation level creates an \gls{aceg} for each \gls{ceg}. 
As a result, the amount of \glspl{aceg} is equal to the amount of \glspl{ceg}.
In comparison to the detailed level that the \glspl{ceg} provides, the \nth{1} aggregation level provides a lower level of detail but allows a more general analysis of the executed process instances.
As an example, \Cref{fig:aggregation_level_1} depicts the constructed \glspl{aceg} from the \gls{ceg} depicted in \Cref{fig:ceg_construction_nodes}.

\begin{figure}[t]
    \centering
    \begin{subfigure}[t]{0.3\textwidth}
        \centering
        \tikzset{>=latex}
        \begin{tikzpicture}[level distance=2.8cm, text width=3cm, align=center, edge from parent/.style={draw,-latex}, font=\footnotesize]
            \node (or1) {Receive \\Purchase Order\\(2)}
            child {
                node (pi1) {Pick Order Items\\(4)} 
                child {
                    node (sh1) {Register \\Shipment\\(2)}
                    edge from parent node[left=-0.6, font=\footnotesize] {1..2 : 1..1\\(4)}
                }
                edge from parent node[left=-0.6, font=\footnotesize] {1..1 : 1..3 \\ (2)}
            };
        \end{tikzpicture}
        \caption{Transformed \gls{aceg} of $ceg_1$ and $ceg_2$.}
        \label{fig:aggregation_level_2_1}
    \end{subfigure}
    \qquad
    \begin{subfigure}[t]{0.3\textwidth}
        \centering
        \tikzset{>=latex}
        \begin{tikzpicture}[level distance=2.8cm, text width=3cm, align=center, edge from parent/.style={draw,-latex}, font=\footnotesize]
            \node (or2) {Receive \\Purchase Order\\(1)}
            child {
                node (pi4) {Pick Order Items\\(1)} 
                child {
                    node (co1) {Register \\Customer Pickup\\(1)}
                    edge from parent node[left=-0.6, font=\footnotesize] {1..1 : 1..1\\(1)}
                }
                edge from parent node[left=-0.6, font=\footnotesize] {1..1 : 1..1\\(1)}
            };
        \end{tikzpicture}
        \caption{Transformed \gls{aceg} of $ceg_3$.}
        \label{fig:aggregation_level_2_3}
    \end{subfigure}
    \caption{\nth{2} aggregation level of the \gls{ceg} depicted in \Cref{fig:ceg_construction_nodes}.}
    \label{fig:aggregation_level_2}
\end{figure}
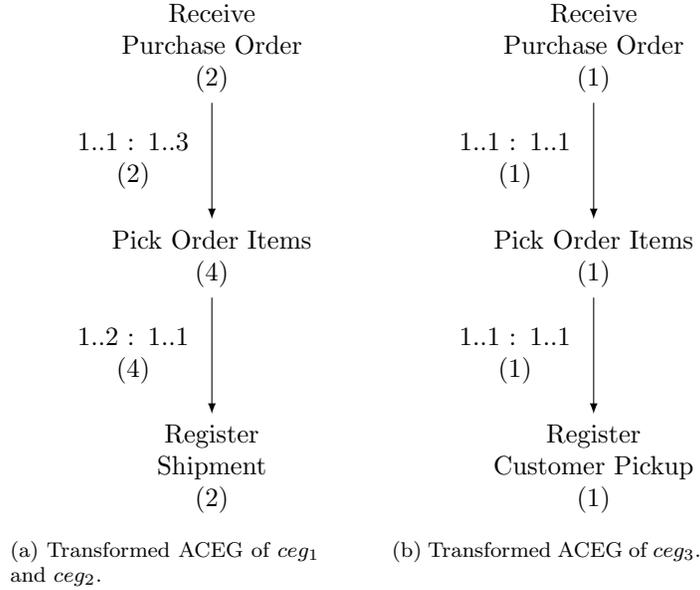

\paragraph{Aggregation Level 2}
The \nth{2} aggregation level creates an \gls{aceg} for all \glspl{ceg} that have the same structure. 
Two \gls{ceg} have the same structure \emph{iff} the event types $ET$, relationships $\leq$, and labels $\lambda$ of the corresponding \gls{aceg}, from the \nth{1} aggregation level, are equal. 
As a result, the amount of \glspl{aceg} is lower or equal to the amount of \glspl{ceg}.
Again, this aggregation level decreases the level of detail in comparison to the \nth{1} aggregation level but provides a higher view of the executed process instances.
As an example, \Cref{fig:aggregation_level_2} depicts the transformed \glspl{aceg} from the \gls{ceg} depicted in \Cref{fig:ceg_construction_nodes}. 
As can be seen, $CEG_1$ and $CEG_2$ are aggregated to one \gls{aceg} since they have the same structure. $CEG_3$, on the other hand, has another structure, since the last event is a \emph{Register Customer Pickup} event instead of a \emph{Register Shipment} event.

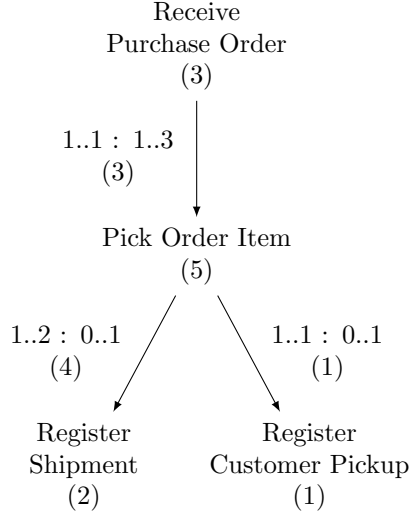
\begin{figure}[t]
    \centering
    \tikzset{>=latex}
    \begin{tikzpicture}[level distance=2.8cm, sibling distance=3cm, text width=3cm, align=center, edge from parent/.style={draw,-latex}, font=\footnotesize]
        \node (or1) {Receive \\Purchase Order\\(3)}
        child {
            node (pi1) {Pick Order Item\\(5)} 
            child {
                node (sh1) {Register \\Shipment\\(2)}
                edge from parent node[left=-0.6, font=\footnotesize] {1..2 : 0..1\\(4)}
            }
            child {
                node (cu1) {Register \\Customer Pickup\\(1)}
                edge from parent node[right=-0.6, font=\footnotesize] {1..1 : 0..1\\(1)}
            }
            edge from parent node[left=-0.6, font=\footnotesize] {1..1 : 1..3 \\ (3)}
        };
    \end{tikzpicture}
    \caption{\nth{3} aggregation level of the \gls{ceg} depicted in \Cref{fig:ceg_construction_nodes}.}
    \label{fig:aggregation_level_3}
\end{figure}

\paragraph{Aggregation Level 3}
The \nth{3} aggregation level creates one \gls{aceg} for the whole $CEG_{DB}$, i.e., all \glspl{ceg}. 
Therefore, this aggregation level is the highest and provides a complete view of the executed process instances.
For instance, \Cref{fig:aggregation_level_3} depicts the transformed \glspl{aceg} from the \gls{ceg} depicted in \Cref{fig:ceg_construction_nodes}.

These three aggregation levels together with the \glspl{ceg} provide the means to analyze the process on the lower instance level as well as on different higher levels, with the process model at the highest level. 

%% file: sections/30_approach/34_graph_analysis.tex
\subsection{Causal Event Graph Analysis}
\label{subsec:graph-analysis}

\noindent
In general, the analysis possibilities provided by the Noreja Approach can be categorized into \emph{Visualisation}, \emph{\glspl{kpi}}, and \emph{Violations}. 
We will discuss these categories and give some examples of how the analysis can be done using the Noreja Approach. 

\subsubsection{Visualisation}
\noindent
The first analysis category of visualisations presents the \glspl{ceg} and \glspl{aceg} to the analyst. 
This provides the analyst with a way to visually inspect the processes on the different detail levels, i.e., from a detailed process instance level provided by the \glspl{ceg} to a high process model level provided by the \glspl{aceg}.
The \glspl{ceg} and \glspl{aceg} can be shown to the analyst as graphs, similar as, e.g., \Cref{fig:ceg_construction_nodes} for the \gls{ceg} level or \Cref{fig:aggregation_level_3} for the \gls{aceg} level.
Moreover, by displaying the \glspl{aceg} from Aggregation Level 2, the analyst can analyze the different process variations as shown in \Cref{fig:aggregation_level_2}.


\subsubsection{\acrlong{kpi}}
\noindent
The second analysis category focuses on the representation of \glspl{kpi} of the observed fragments.
Different \glspl{kpi} can be calculated, among them are, for instance: 
\begin{itemize}
    \item \emph{Minimum, Average, and Maximum Cycle Time of an Event Type}: The calculation of this \gls{kpi} is shown in \Cref{alg:min_avg_max_event_duration}.
    The algorithm takes as input a set of \glspl{ceg} and an event type. 
    The algorithm, then iterates through the \glspl{ceg} and calculates the minimum, average, and maximum event type cycle time of the requested event type.
    \item \emph{Minimum, Average, and Maximum \gls{ceg} Cycle Time}: The calculation of this \gls{kpi} is shown in \Cref{alg:min_avg_max_case_duration}. 
    This algorithm takes as input a set of \glspl{ceg}. 
    The algorithm then iterates through the set of \glspl{ceg} and searches for the 
 event with the smallest timestamp, i.e., $min_{ceg} = \min(\{t(e)~|~e \in E_{ceg}\})$, and the event with the 
    biggest timestamp, i.e., $max_{ceg} = \max(\{t(e)~|~e \in E_{ceg}\})$. 
    The found $min_{ceg}$ and $max_{ceg}$ timestamps are then used to calculate the \gls{ceg} cycle time which is added to a set of cycle time.
    Eventually, the collected cycle time are used to calculate the minimum, the average, and the maximum \gls{ceg} cycle time.
    \item \emph{Minimum, Average, and Maximum Fragment Cycle Time}: The calculation of this \gls{kpi} is similar to the one for the \gls{ceg} cycle time, but on a fragment level. Therefore, \Cref{alg:min_avg_max_case_duration} gets as input a set of fragments instead of a set of \glspl{ceg}.
    \item \emph{Distribution End Event Types}: This \gls{kpi} shows the distribution of end event types of the observed paths. For this \gls{kpi}, the \nth{3} aggregation level \gls{aceg} is used. A set of all end event types, together with the corresponding quantity, is created as follows: $\mathbb{E}=\{et\in ET ~|~\post{et} = \emptyset\}$. This set can then be used to calculate the absolute and relative distribution of end events. 
    \item \emph{Batching Event Type Distribution}: To calculate the distribution of batching event types, we first get a list of batching events by $\mathbb{B} = \{e\in E~|~batch(e)=1\}$. Afterwards the absolute and relative distribution can be calculated by using the event types of the batching events: $\mathbb{BT} = \{e^{\tau}~|~e \in \mathbb{B}\}$. 
\end{itemize}
\begin{algorithm}[t]
    \caption{Get Minimum, Average, and Maximum Event Type Cycle Time}\label{alg:min_avg_max_event_duration}
    \begin{algorithmic}[1]
        \Require {$C = \{ CEG_1, CEG_2, ..., CEG_n \}; an \; event \; type \;et$}
        \Ensure{$min, avg, max$ event type cycle time}
        \State $cycleTime = \emptyset$
        \ForAll{$CEG= \langle E, \leq, \lambda \rangle \in C$}
            \ForAll{$e \in E$}
                \If{$e^{\tau} = et$}
                    \State $cycleTime = cycleTime \cup \mathrm{cyc}(e)$;
                \EndIf
            \EndFor   
        \EndFor     
        \State $min = min(cycleTime)$;
        \State $max = max(cycleTime)$;
        \State $avg = avg(cycleTime)$;
        \State \Return $min, avg, max$;
    \end{algorithmic}
\end{algorithm}
\begin{algorithm}[t]
    \caption{Get Minimum, Average, and Maximum CEG cycle time}\label{alg:min_avg_max_case_duration}
    \begin{algorithmic}[1]
        \Require {$C = \{ CEG_1, CEG_2, ..., CEG_n \}$}
        \Ensure{$min, avg, max$ fragment cycle time}
        \State $cycleTime = \emptyset$
        \ForAll{$CEG=\langle E, \leq, \lambda \rangle \in C$}
            \State $min_{CEG} = \min(\{t(e)~|~e \in E \})$;
            \State $max_{CEG} = \max(\{t(e)~|~e \in E \})$;
            \State $cycleTime_{CEG} = max_{CEG} - min_{CEG}$;
            \State $cycleTime = cycleTime \cup cycleTime_{CEG}$;
        \EndFor     
        \State $min = min(cycleTime)$;
        \State $max = max(cycleTime)$;
        \State $avg = avg(cycleTime)$;
        \State \Return $min, avg, max$;
    \end{algorithmic}
\end{algorithm}

\subsubsection{Violations}
\label{subsec:violations}
\noindent
The third analysis category is about the analysis of violations.
The focus here is the temporal order of events as observed in comparison to what is expected according to the \gls{cpt}.
Such temporal violations can be detected by comparing the timestamps between two events, $e_a, e_b \in E$, for which there exists a causal relationship $e_a \leq e_b$. 
A causal relationship $e_a \leq e_b$ defines that an event $e_a$ causes an event $e_b$, as a consequence the timestamp $t(e_a)$ should be before $t(e_b)$. 
If this is not the case, the causal relationship is violated. 
With a given set of \gls{ceg}, a set with all the pairs that violated the temporal order can be created as follows: $\mathbb{V}_o=\{(e_a,e_b) \in \leq | t(e_a) < t(e_b)$\}.

%% file: sections/40_evaluation/40_evaluation.tex
\section{Evaluation}
\label{sec:evaluation}

\noindent 
In the following, we evaluate the Noreja Approach by comparing it to state-of-the-art process discovery algorithms. 
To this end, we first discuss the dataset that we use and our prototypical implementation. 
Then, we compare the solutions qualitatively and quantitatively.   

\subsection{Evaluation Case}
\noindent 
We use a real-world dataset from our industry partner, an European food production company, which captures the events of the order-to-cash process. 
The process is triggered by supermarket orders (a). 
These orders are broken down into a list of item suborders (b). Each of these items is separately picked from the warehouse (c). Then commissioned (and packaged) (d) and sent as one or multiple deliveries (e). 
Finally, an invoice is sent to the customer~(f). 
Some of these steps also contain substeps. 
Furthermore, each step of an order can also be batched together with another order, e.g., several orders might be batched in one delivery. 

\begin{figure}[t]
	\centering
	\includegraphics[width=1.0\columnwidth]{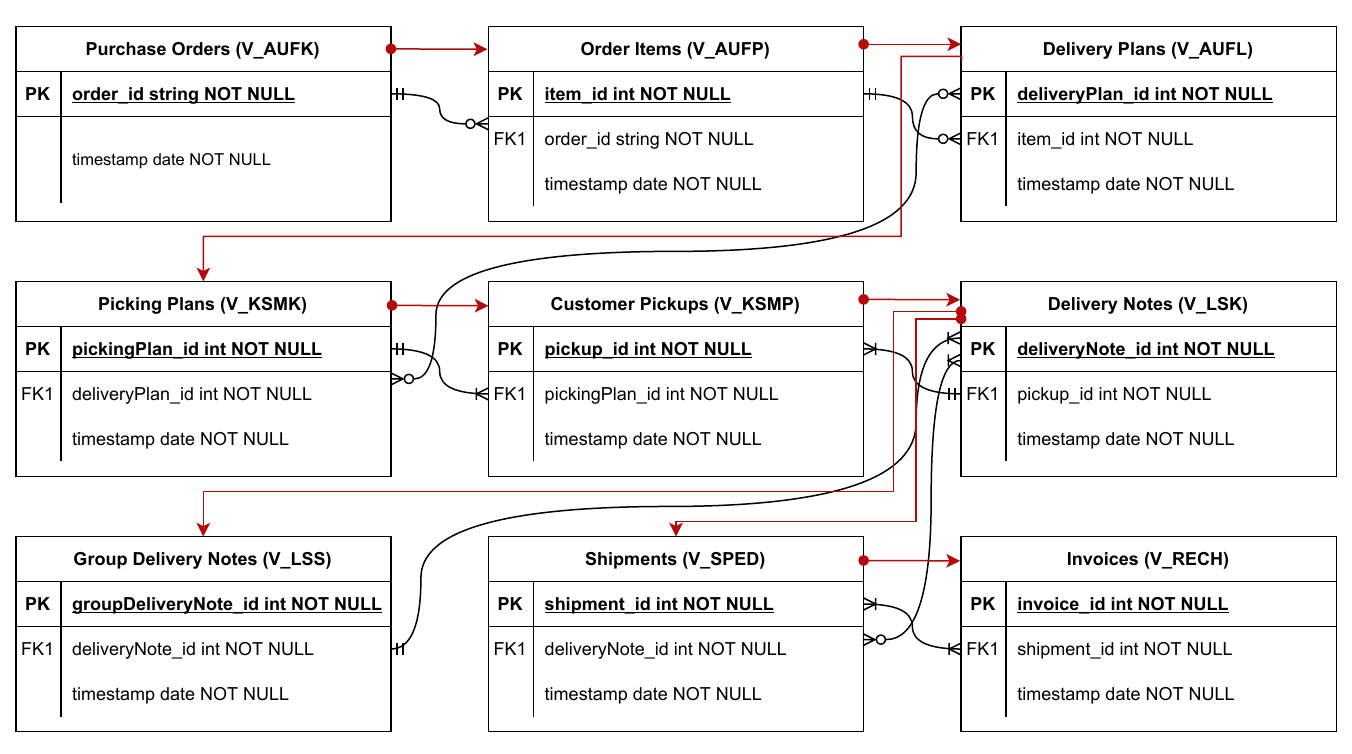}
	\caption{ER diagram of source data tables and \gls{cpt} of the Order-to-Cash process (in red). \emph{Note:} We link the following event labels to the tables: V\_AUFK: Receive Order, V\_AUFP: Extract Order Items, V\_AUFL: Prepare Order Delivery, V\_KSMK: Plan Order Item Picking, V\_KSMP: Pick Order Item, V\_LSK: Create Delivery Note, V\_LSS: Generate Group Delivery Note, V\_SPED: Deliver Order, V\_RECH: Post Invoice.}
	\label{fig:er_evaluation}
\end{figure} 

%
The company uses an ERP system built on a Microsoft SQL Server, which serves as the data source for our evaluation. We used the following catalog of tables: $Cat=\{V\_AUFK, V\_AUFP, V\_AUFL, V\_KSMK, V\_KSMP,\\ V\_LSK, V\_LSS, V\_SPED, V\_RECH\}$. Figure \ref{fig:er_evaluation} depicts the ER-diagram for $Cat$ and also shows the \gls{cpt} describing the partial order assumed among the relations in $Cat$ (in red). This partial order was built based on the foreign key relations and the causal knowledge of the analyst. In total, the dataset contains 70,000 orders with more than 8,500,000 events. 

For each event, the ERP system stores, besides different resource properties, the start timestamp that marks the start of an event, and the last update timestamp that marks the end of an event. 
For the evaluation, we use the raw event data without any data cleaning beforehand. 
Note that publicly available dataset of the BPI Challenges are flat event logs. As our approach leverages external knowledge about causal structures as available in a relational database, we could not use these public datasets.

\subsection{Prototype}
\noindent 
Our prototype is implemented as a component of the commercial Noreja tool as a Java microservice-based application, using Spring Boot (vers. 2.4.4), and a TypeScript based Angular (vers. 12.2.10) web \gls{ui}. 
To store and query the \glspl{ceg} and \glspl{aceg}, we use the graph database Neo4j (vers.~4.3.0) and Cypher as the Neo4j query language. The current prototype can read event data from Microsoft SQL Server and Oracle~DB. 

\begin{figure}
    \centering
    \includegraphics[width=1\columnwidth]{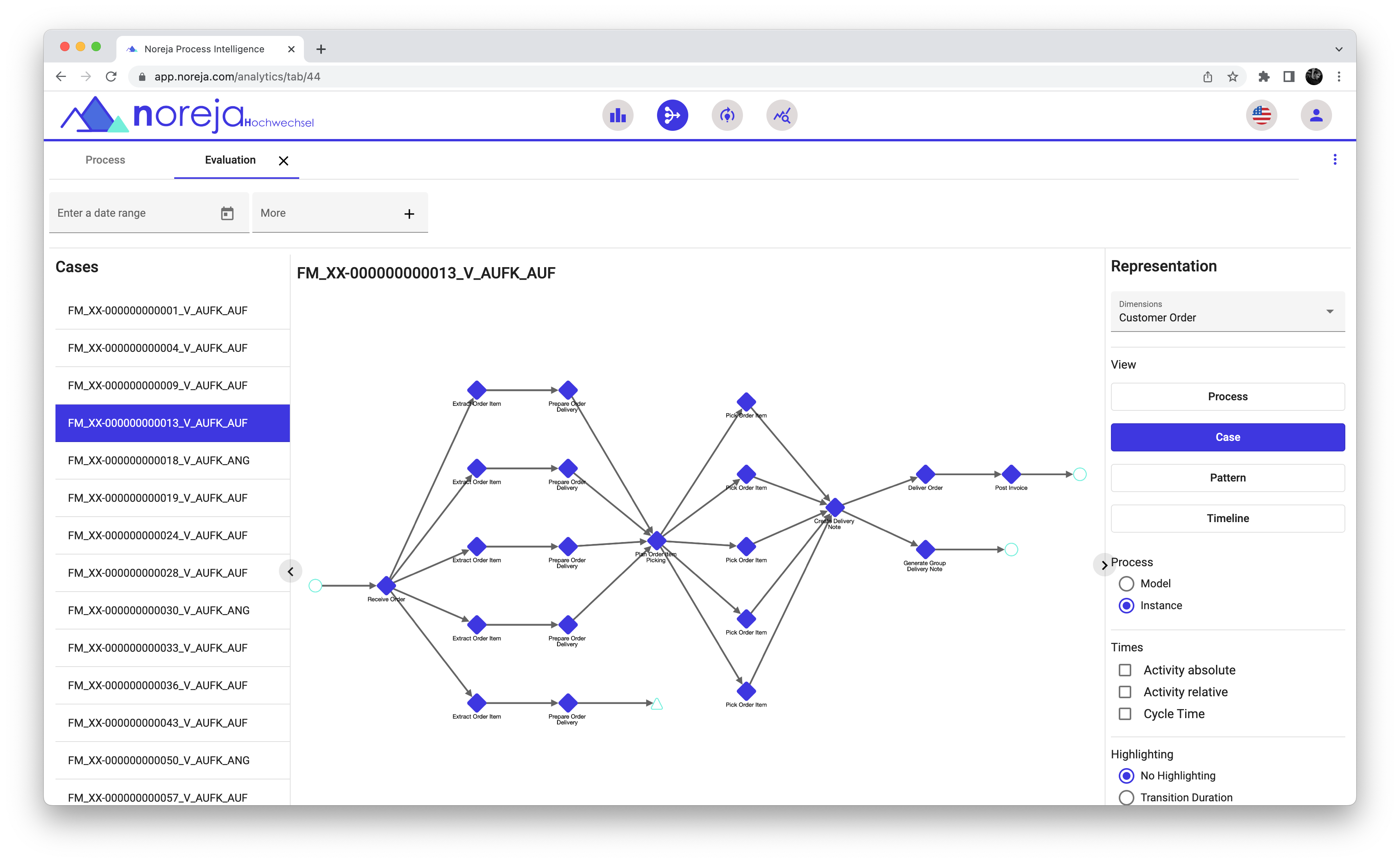}
    \caption{\gls{ceg} created with the Noreja Approach.}
    \label{fig:noreja-graph-ceg}
\end{figure}

\begin{figure}
    \centering
    \includegraphics[width=1\columnwidth]{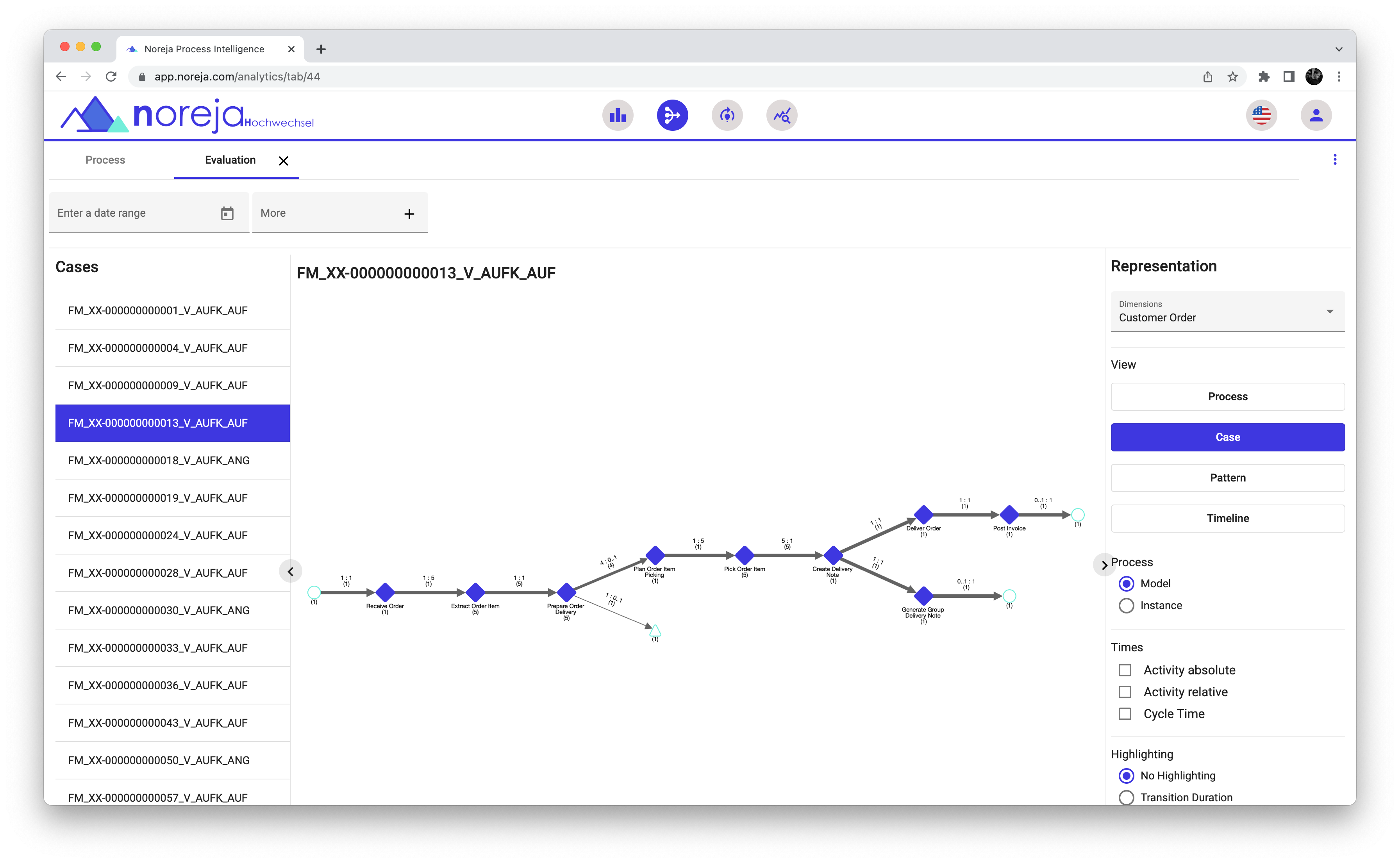}
    \caption{\nth{1} aggregation level \gls{aceg} created with the Noreja Approach.}
    \label{fig:noreja-graph-aceg-level-1}
\end{figure}

\begin{figure}
    \centering
    \includegraphics[width=1\columnwidth]{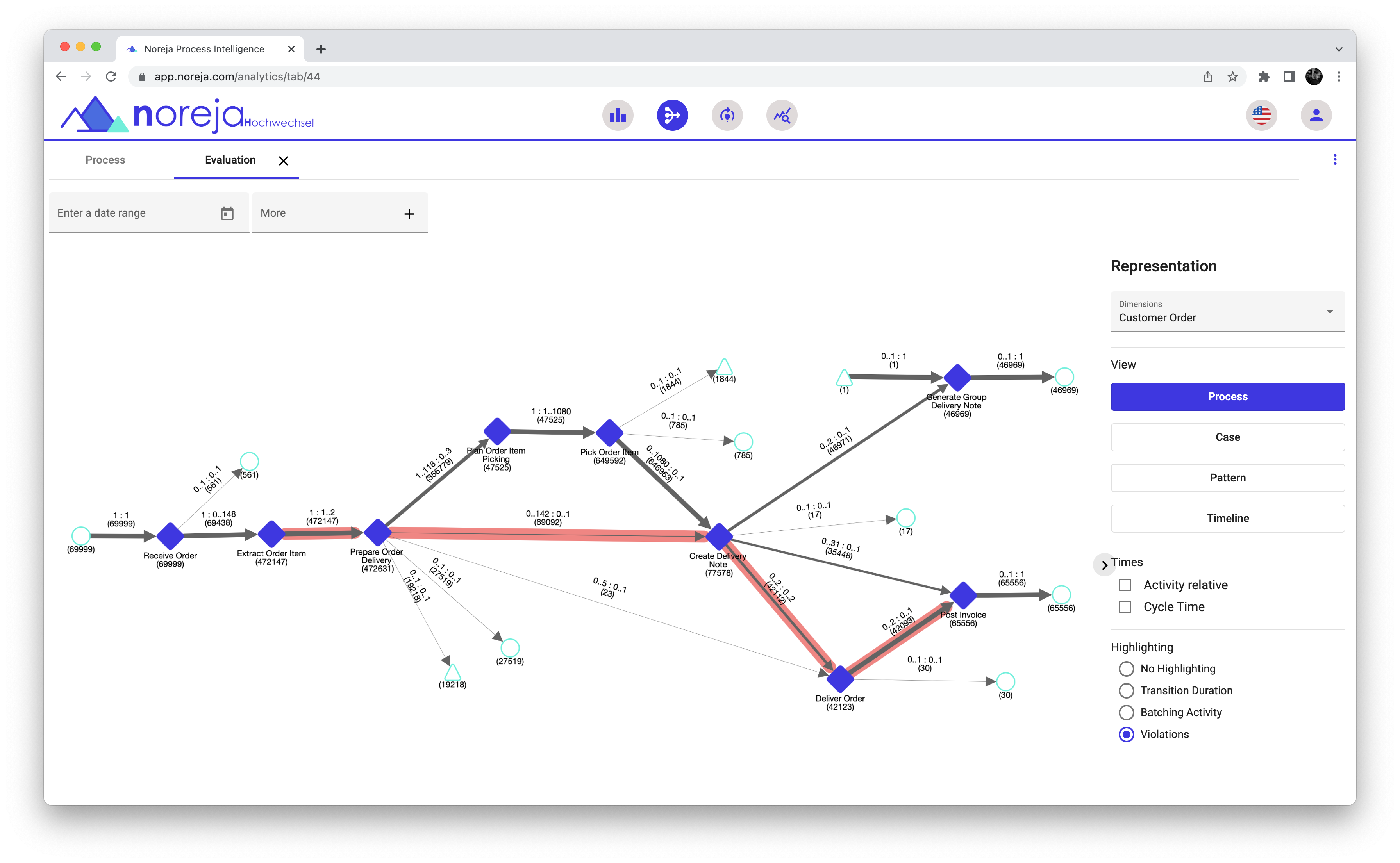}
    \caption{With violation highlighting.}
    \label{fig:noreja-graph-aceg-level-3-violations}
\end{figure}

\Cref{fig:noreja-graph-ceg} to \Cref{fig:noreja-graph-aceg-level-3-violations} present three screenshots of the \gls{ui}. 
\Cref{fig:noreja-graph-ceg} shows a single \gls{ceg}, which was selected from the list of available \glspl{ceg} on the left side. 
Events are visualized as blue rhombuses, called \emph{diamonds}. 
Cyan-colored circles and triangles are used to mark the start and end of a process. 
A triangle marks a start or end of a path for which there is a parallel path that continous, e.g., the path at the bottom after \textit{Prepare Order Delivery} in \Cref{fig:noreja-graph-ceg}, these triangle markers are called \emph{intermediate start} or \emph{end}. 
A circle marks the start or end of the process for which there is no parallel continuing path. 
\Cref{fig:noreja-graph-aceg-level-1} shows the \nth{1} aggregation level \gls{aceg} of the \gls{ceg} shown in \Cref{fig:noreja-graph-ceg}. 
As discussed in \Cref{subsec:graph-aggregation}, the relationship and node quantity are shown in parenthesis and the relationship cardinality is shown in the format \{$\rtimes_{min}$..$\rtimes_{max}$ : $\ltimes_{min}$..$\ltimes_{max}$\}. 
\Cref{fig:noreja-graph-aceg-level-3-violations} shows the \nth{3} aggregation level \gls{aceg} with a highlighting of relationships that contains a temporal violation (in red).

\subsection{Qualitative Analysis}
\begin{figure}
    \centering
    \includegraphics[width=1.0\columnwidth]{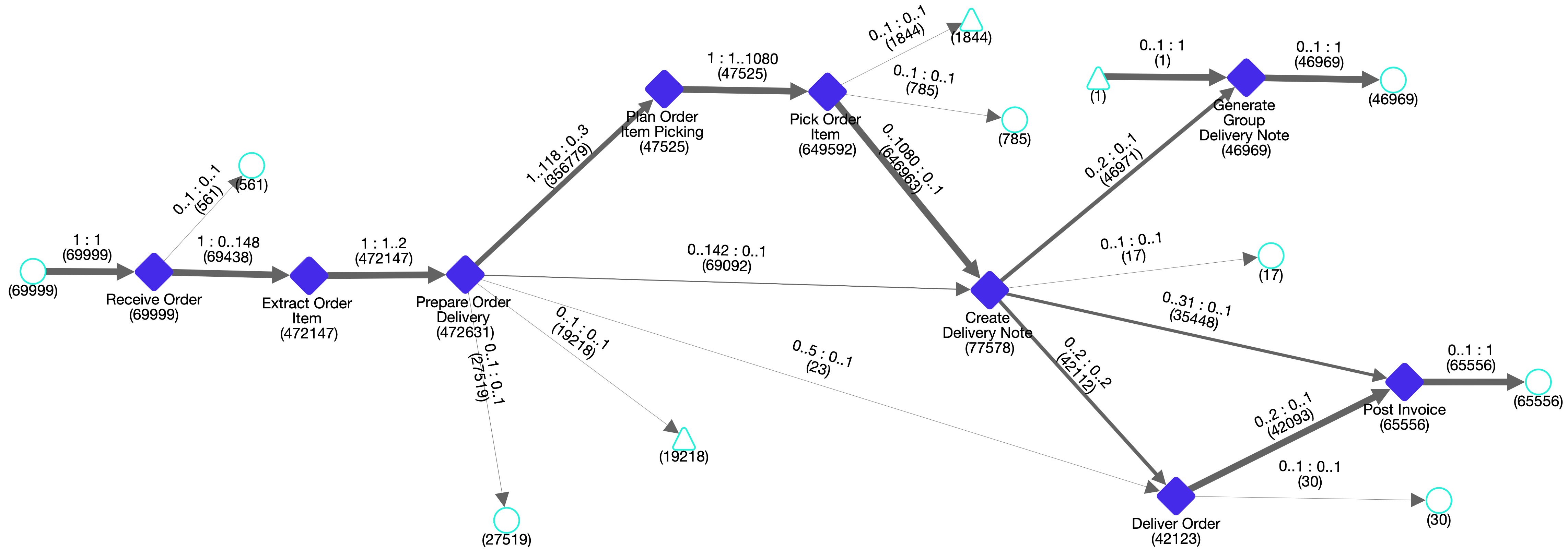}
    \caption{\nth{3} aggregation level \gls{aceg} created with the Noreja Approach.}
    \label{fig:noreja-discovered-aceg-level-3}
\end{figure}

\begin{figure}
    \centering
    \includegraphics[width=1.0\columnwidth]{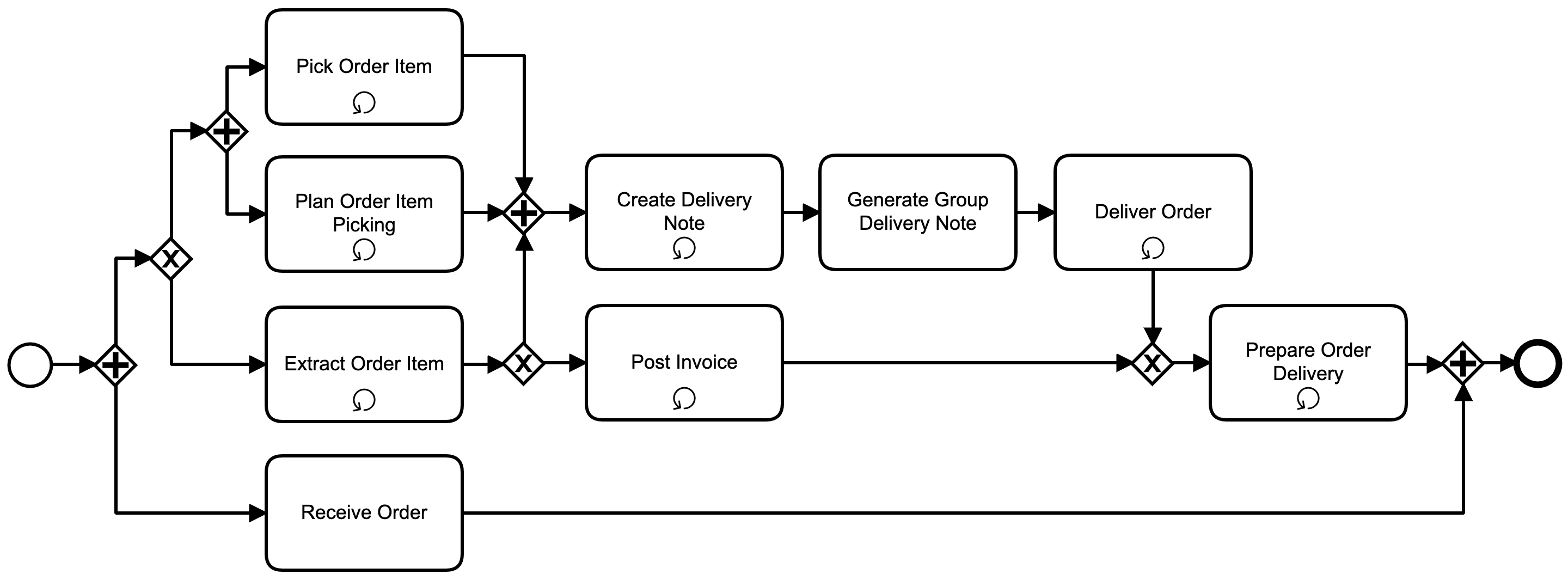}
    \caption{\gls{bpmn} created by Split Miner with a concurrency threshold of 0.2. \emph{Note:} The activities got re-arranged for a clearer view. }
    \label{fig:splitminer-graph-1}
\end{figure}

\begin{figure}
    \centering
    \includegraphics[width=0.65\columnwidth]{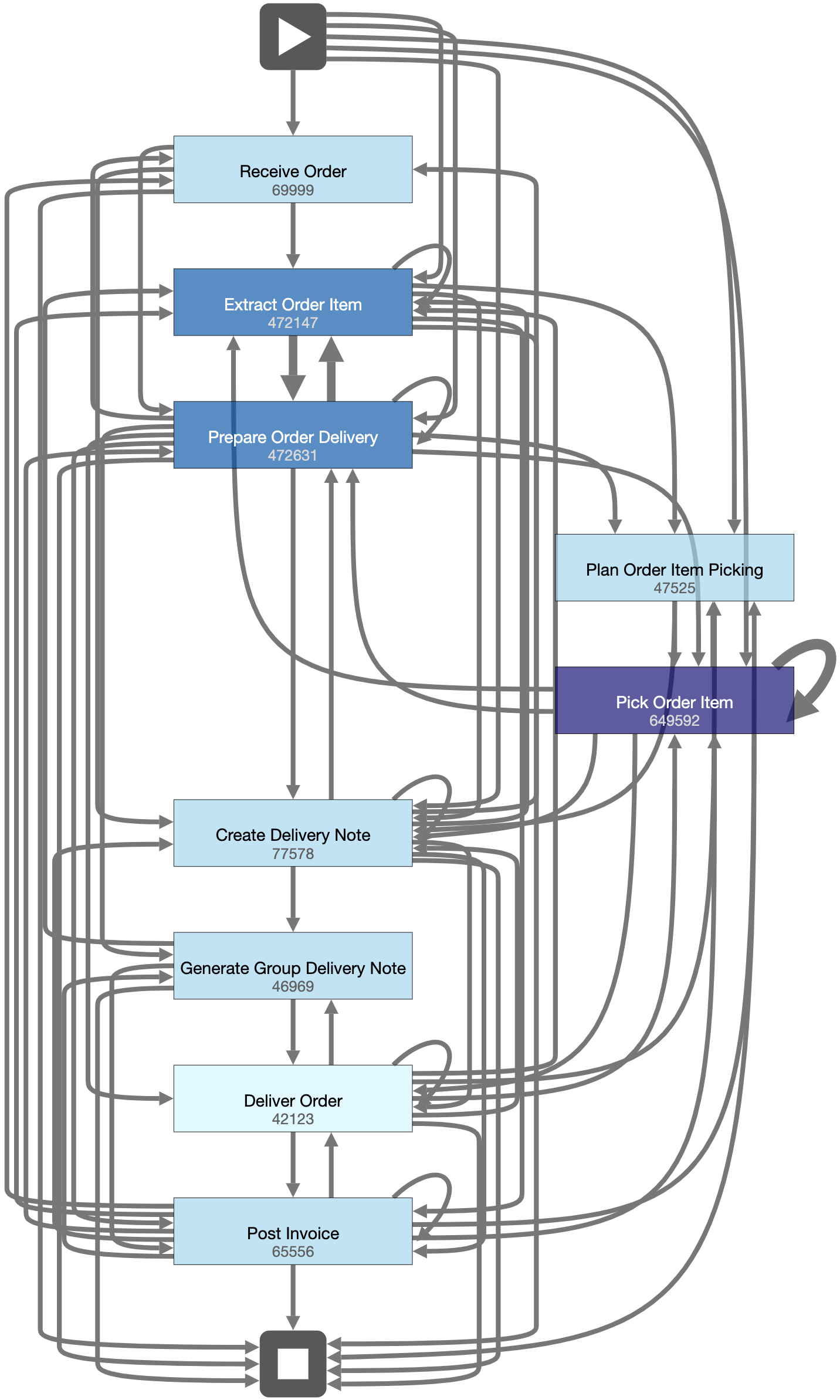}
    \caption{\gls{dfg} created with PMTK.}
    \label{fig:pmtk-graph}
\end{figure}

\noindent
To evaluate the presented Noreja Approach, we first compare the structure it learned, i.e., the CEG with the structure mined from state-of-the-art approaches. 
More specifically, we compare the \nth{3} aggregation level \gls{aceg} produced with the Noreja Approach (cf. \Cref{fig:noreja-discovered-aceg-level-3}) with the \gls{bpmn} 2.0 model created by Split Miner 2.0 \citep{DBLP:conf/icpm/AugustoDR20} (cf. \Cref{fig:splitminer-graph-1}) and a \gls{dfg} created by PMTK \citep{pmtk}, which is a \gls{ui} for PM4Py \citep{DBLP:journals/corr/abs-1905-06169} (cf. \Cref{fig:pmtk-graph}). 
%

As can be seen in \Cref{fig:pmtk-graph}, which shows the \gls{dfg} discovered by PM4Py, the \gls{dfg} contains several spurious relationships that are not expected considering the causality among the relations (see Figure \ref{fig:er_evaluation}). 
For instance, most events contain self-loops as a byproduct of the 1:N, N:1, and N:M relationships in the database schema. 
They stem from separate execution sub-streams that create spurious direct-follows observations. 
However, this kind of self-loops are not possible in the actual database schema of the ERP system. 
Another example of a spurious relationship is the relationship from the \textit{Pick Order Item} event to the \textit{Prepare Order Delivery} event. 
This relationship suggests loops of the kind $\langle$ \emph{Prepare Order Delivery, Plan Order Item Picking, Pick Order Item, Prepare Order Delivery, \dots} $\rangle$, which contradict the causal relationships of the event sequences (cf. \Cref{fig:er_evaluation}) and the database schema. 
These kind of loops are not present in the dataset, but a by-product of creating the process model without considering the causal relationships. 
In comparison, the \gls{aceg} created by the Noreja Approach, depicted in \Cref{fig:noreja-discovered-aceg-level-3}, does not contain these spurious relationships, i.e., the \gls{aceg} does not contain self-loops and does not contain the relationship between \textit{Pick Order Item} and \textit{Prepare Order Delivery}. 
The reason for this is that the Noreja Approach uses the causal knowledge defined by the process analyst shown in \Cref{fig:er_evaluation}.

When we compare the discovered \gls{bpmn} from Split Miner 2.0 (cf. \Cref{fig:splitminer-graph-1}) with the expected order of activities (cf. \Cref{fig:er_evaluation}), we can see that some activity orders got discovered correctly. 
For instance, the \gls{bpmn} defines that before \emph{Delivery Order} there has to be a \emph{Create Deliver Node} event.
Another similarity is that each process instance has a \emph{Receive Order} event. 
However, there are also differences. 
According to the causality among the relations (cf. \Cref{fig:er_evaluation}), the event \emph{Prepare Order Delivery} is executed before the event \emph{Deliver Order}. 
However, in the discovered process model the order is the other way around, which violates the causality between these events.
Another difference is that according to the expected process, the events \emph{Plan Order Item Picking} and \emph{Pick Order Item} are executed in sequence. 
In the discovered process model, they are executed in parallel. 
This can be attributed to how the original ERP system creates the \textit{Extract Order Item} and \textit{Prepare Order Delivery} events. 
After a new order is stored, the events \textit{Extract Order Item} and \textit{Prepare Order Delivery} are automatically created and stored shortly after another. 
In some cases, the automatic creation and storage of the \textit{Prepare Order Delivery} events are faster (in the millisecond range) than the creation and storage of the events \textit{Extract Order Item}.
Similar to the \gls{dfg} created by PM4Py, most of the activities contain self-loops. 
These self-loops are again a by-product of the database schemas 1:N, N:1, and N:M relationships. 
When we compare the \gls{bpmn} with the \gls{aceg} created with the Noreja Approach, we can see that by using the causal knowledge from the \gls{cpt}, the order of the activities is preserved. Moreover, by using the \gls{cpt}, the Noreja Approach overcomes the problem of self-loops. 

\subsection{Quantitative Analysis}
\label{subsec:metrics-evaluation}
\noindent 
Next, we compare the conformance of the \gls{dfg} discovered by PM4Py and the \gls{aceg} created with the Noreja Approach by comparing them with the causality relations defined (cf. \Cref{fig:er_evaluation}). 
As a metric, we calculate for each causal relationship, i.e., for each pair in $\prec$, the ratio between the quantity of the expected relations to the unexpected one. 
For instance, when we consider the defined causality relationship \textit{Receive Order} $\prec$ \textit{Extract Order Item} and calculate the ratio for the \gls{aceg} shown in \Cref{fig:noreja-discovered-aceg-level-3} we get 69,438 / 561 (0.81\%) since we have a quantity of 69,438 from \textit{Receive Order} to \textit{Extract Order Item} (the expected relationship) and a quantity of 561 from \textit{Receive Order} to the end (the unexpected relationship). 
We perform this calculation for the \gls{dfg} created by PM4Py and the \gls{aceg} created with the Noreja Approach. 
We do not consider the \gls{bpmn} discovered by Split Miner since it does not show quantities. 
The resulting ratios for the \gls{dfg} created by PM4Py can be found in \Cref{tab:pm4py_expected_unexpected_quantity} and for the \gls{aceg} in \Cref{tab:noreja_expected_unexpected_quantity}. 
In addition, \Cref{tab:quantity_withoutviolation_violations} shows for the \gls{aceg} the ratio between the quantity of relationships without temporal violations (as defined in \Cref{subsec:violations}) to the quantities of relationships with a temporal violation. 

As can be seen in the total column of \Cref{tab:pm4py_expected_unexpected_quantity} and \Cref{tab:noreja_expected_unexpected_quantity}, most unexpected relationships quantities are higher in the case of the \gls{dfg} than for the \gls{aceg}. 
For instance, the relationship between \textit{Pick Order Item} and \textit{Create Delivery Note} has a ratio of 46,124 / 603,468 (1,308.36\%) for the \gls{dfg} and a ratio of 646,963 / 2,629 (0.41\%) for the \gls{aceg}. 
In the case of the \gls{dfg}, the huge amount of unexpected relationships stems from self-loops (due to 1:N relationships in the database schema), but also from relationships from \textit{Pick Order Item} to activities at the beginning of the process e.g., \textit{Extract Order Item}. 
In the case of the latter, these unexpected relationships stem from situations in which the order got updated e.g., a product was added at a later point in time. 

By considering the causal knowledge defined in the \gls{cpt}, the Noreja Approach does not suffer from the 1:N relationship problem and thus has a smaller amount of unexpected relationships. 
The situation of unexpected relationships when an order got updated, does not add an unexpected relationship to the \gls{aceg}, but it results in a temporal violation of a relationship before the current event, e.g., before \textit{Pick Order Item}. 
For instance, as can be seen in \Cref{tab:quantity_withoutviolation_violations} the relationship from \textit{Prepare Order Delivery} to \textit{Create Delivery Note} contains multiple violations, suggesting that some \textit{Prepare Order Delivery} events happened after \textit{Create Delivery Note}. 

Another difference can be seen for the ratio of the relationship \textit{Post Invoice} to the end. 
In case of PM4Py's \gls{dfg}, we have a ratio of 63,758 / 1,798 (2.82\%) and in case of Noreja's \gls{aceg} we have 65556 / 0 (0.00\%). 
As can be seen in \Cref{fig:pmtk-graph} the event \textit{Post Invoice} has outgoing relationships to all other events in the case of the \gls{dfg}, despite the fact that this event should be the last one. 
Again the reasons for the number of unexpected relationships are updates of the orders after the \textit{Post Invoice} event was already created. 

To quantify the resulting conformance to the defined CPT, we use the following metric based on the unexpected relationships and temporal violations: 
\begin{itemize}
    \item For PM4Py's \gls{dfg}, we calculate the cumulative quantity of all unexpected relationships from \Cref{tab:pm4py_expected_unexpected_quantity}.
    \item For Noreja's \gls{aceg}, we also calculate the cumulative quantity of all unexpected relationships from \Cref{tab:noreja_expected_unexpected_quantity}. Furthermore, we add the quantity of all relationships with a temporal violation according to \Cref{tab:quantity_withoutviolation_violations}, when they are on an expected relationship. 
\end{itemize}

By using this metric, we obtain a score of 1,153,710 for the \gls{dfg} discovered by PM4Py and a score of 165,124 for the \gls{aceg} created by the Noreja Approach. 
While the absolute numbers are not directly expressive, the numbers relative to each other are. 
The significantly smaller score of the \gls{aceg} suggests a much simpler process with higher conformance to the defined causality. 
It has to be noted that in the case of the \gls{dfg}, the main part of the high score is due to the self-loops in the case of the \textit{Pick Order Item} event and the relationship between the \textit{Extract Order Item} and \textit{Prepare Order Delivery} events. 
As discussed above, the automatic creation and storage of the \textit{Prepare Order Delivery} events are sometimes faster than the creation and storage of the events \textit{Extract Order Item}, which leads to alternating executions. 
However, when the unexpected relationships that start with \textit{Extract Order Item} and \textit{Prepare Order Delivery} events are ignored, the cumulative score of the \gls{dfg} is 687,053, which is still much higher than the one for the \gls{aceg}. 

\subsection{Discussion}
\noindent
%
In the following, we reflect on how of predefined requirements in \Cref{subsec:Requirements} are covered by the Noreja Approach.

The first requirement, \emph{Input Data representing Causal Event Structure}, is about the input data that is used for the \gls{pm} technique. As discussed in \Cref{subsec:graph-construction}, our approach uses as a data input source the relational database to keep and represent the causality of the events. We hereby concur with works of \citet{DBLP:journals/tsc/LuNWF15}, \citet{DBLP:conf/apn/Fahland19}, and \citet{10.1007/978-3-030-46633-6_2} who also interpret the underlying cardinalities of relational databases as one potential source of causal knowledge. This stands in contrast to the usage of event logs that mainly discard this kind of causality knowledge due to flattening \citep{Aalst2011OnTR}. Nevertheless, we would like to note that other potential input data sources (e.g., document databases, graph databases, key-value databases, or other NoSQL databases) should not be generally excluded here.

Subsequently, the second requirement, \emph{External Knowledge about Causal Event Structure}, is concerned with the integration or enrichment of \textit{real-world} domain knowledge to observe the causality between the events correctly. \citet{8456350} also resort to this type of information source, particularly for reconstructing and evaluating temporal and sequential orders but they are not using a structured procedure to add this kind of information. By contrast, and in the case of the presented \gls{cpm} approach, this domain knowledge is incorporated by the \gls{cpt}. During the transformation of the source input data, which originates from the relational database, this \gls{cpt} is used as an information source for the causality between the events. As can be seen in the evaluation, by using the \gls{cpt}, the Noreja Approach is able to create simpler process models than those created by state-of-the-art approaches. Additionally, it provides the opportunity to change the provided external knowledge flexibly ex-post (e.g., in case the external environment conditions change).

While the first two requirements are concerned with transforming the input data to the graph structure, the \nth{3} and \nth{4} requirements are concerned with the analysis of the data itself. As defined in \Cref{subsec:Requirements}, the \nth{3} requirement, \emph{Event Cardinalities}, is about the importance of correctly calculating the cardinalities of events in order to cope with the challenge of convergence versus divergence, also called batching behavior. In fact, this challenge has been addressed by a multitude of papers \citep{DBLP:journals/tsc/LuNWF15,DBLP:conf/caise/LiMCA18,10.1007/978-3-030-46633-6_2}. Nevertheless, most of the approaches do not consider cardinalities under the constraint of causality. As our evaluation shows, by leveraging causal knowledge, the Noreja Approach does not suffer from the problems stemming from the 1:N, N:1, and N:M relationships in the database schema. This avoids spurious self-loops and complex structures and provides the means to accurately calculate the cardinalities of the events as well as for the relationships which is mentioned by \citet{Aalst2011OnTR} as one of the most fundamental flaws of common \gls{pm} approaches.

The \nth{4} requirement, \emph{Aggregation of Causal Event Structures}, deals with the question of how an analyst can be supported by providing the process in different levels of detail. 
As discussed in \Cref{subsec:graph-aggregation} and shown in the evaluation, our approach fulfills this requirement by providing each process instance as a \gls{ceg}, i.e., the most detail level, and different aggregations of them as \glspl{aceg}. Other approaches also recognize the rationale for more diverse analysis methods, e.g., by handling multi-instantiated sub-process models  \citep{DBLP:conf/sac/WeberFMS15}, representing it into a process cube \citep{DBLP:journals/corr/abs-2103-07184}, or synthesizing Petri net structures \citep{DBLP:conf/apn/DumasG15}. Somehow, at higher aggregated levels, this often leads to the loss of the causal relationships or the blurring of cardinalities.

%% file: sections/50_discussion.tex



%% file: sections/60_conclusion.tex
\section{Conclusion}
\label{sec:conclusion}
\noindent 
Recent \gls{pm} research has spent considerable effort on developing and improving automatic process discovery algorithms. 
The outcome of these efforts are algorithms with high precision and recall but also complex Spaghetti models. 
In this paper, we developed a novel approach to \glsfirst{cpm} that is based on knowledge about causal relations. 
We presented the Noreja Approach that utilizes the causal knowledge defined in a \glsfirst{cpt} to create a \glsfirst{ceg}. 
Subsequently, different aggregation levels of these \glspl{ceg} are created in form of \glsfirst{aceg}. 
The Noreja approach then exploits these structures to offer multiple operations to analyze the process on different levels of aggregation and from various perspectives. 

Our evaluation demonstrates the benefits of \gls{cpm} and the consideration of causal knowledge during process discovery. 
The evaluation shows that the presented Noreja Approach creates less complex process models than the compared approaches.
One of the main reasons for this is the fact that the Noreja Approach does not struggle with the problem that is inherent with the 1:N, N:1, and N:M relations of database schema.
A second reason for less complex models is attributable to how the Noreja Approach handles situations like later updates of a process instance.
Instead of resulting in additional relationships, which complicate the process model, such situations result in temporal violations.

We are currently working on various extensions of the \gls{cpm} and the Noreja Approach. First, we plan to analyze and develop different kinds of additional visualization, filtering, and aggregation methods. 
Second, we investigate support for drift detection for the process by detecting drifts in the \glspl{ceg} and \glspl{aceg} using machine learning. 
Third, since the transformed \glspl{ceg} and \glspl{aceg} are graph structures, different analysis approaches from graph theory can be used.
For the area of process analysis, especially analysis techniques like \emph{Link Prediction} or \emph{Graph Embeddings} can be beneficial. 
For instance, predicting the next executed activity of an unfinished process or predicting the remaining execution duration. Alternatively, using graph embeddings e.g., Node2Vec~\cite{DBLP:conf/kdd/GroverL16} or Graph2Vec~\cite{DBLP:journals/corr/NarayananCVCLJ17} provides the means to perform advanced levels of clustering. 

%% file: sections/40_evaluation/expected_unexpected_table_pm4py.tex
\renewcommand{\arraystretch}{2.3}
\begin{adjustbox}{angle=90, center, caption={Ratio between the Expected Relationships Quantity and the Unexpected Relationships Quantity (Expected / Unexpected) of the Process Model Discoverd by PM4Py.\label{tab:pm4py_expected_unexpected_quantity}}, nofloat=table}
    \centering
    \tiny
    \begin{tabular}{p{0.05cm}p{1cm}p{0.6cm}p{0.8cm}p{0.9cm}p{1cm}p{1cm}p{0.8cm}p{1cm}p{1cm}p{0.8cm}p{0.8cm}p{0.8cm}p{1.4cm}}
        \toprule
            &         & \multicolumn{12}{c}{Target of Relationship}                                                                                                                                     \\
            &         & Start & Receive Order      & Extract Order Item      & Prepare Order Delivery        & Plan Order Item Picking        & Pick Order Item   & Create Delivery Note         & Generate Group Delivery Note        & Deliver Order     & Post Invoice      & End          & Total (\%)          \\
        \midrule
        \multirow{17}{*}{\STAB{\rotatebox[origin=c]{90}{Source of Relationship}}}
            & Start   & 0 / 0 & 68,836 / 1,163 & 0 / 0        & 0 / 0          & 0 / 0          & 0 / 0     & 0 / 0        & 0 / 0       & 0 / 0     & 0 / 0      & 0 / 0      & 68,836 / 1,163 (1.69\%)  \\
            & Receive Order & 0 / 0 & 0 / 0    & 68,103 / 1,896 & 0 / 0        & 0 / 0        & 0 / 0   & 0 / 0        & 0 / 0       & 0 / 0     & 0 / 0      & 0 / 0      & 68,103 / 1,896 (2.78\%)   \\
            & Extract Order Item & 0 / 0 & 0 / 0    & 0 / 0      & 433,119 / 39,023 & 0 / 0        & 0 / 0   & 0 / 0        & 0 / 0       & 0 / 0     & 0 / 0      & 0 / 0      & 433,119 / 39,023 (9.01\%) \\
            & Prepare Order Delivery & 0 / 0 & 0 / 0    & 0 / 0      & 0 / 0        & 44,997 / 427,634 & 0 / 0   & 0 / 0        & 0 / 0       & 0 / 0     & 0 / 0      & 0 / 0      & 44,997 / 427,634 (950.36\%) \\
            & Plan Order Item Picking & 0 / 0 & 0 / 0    & 0 / 0      & 0 / 0        & 0 / 0        & 47,524 / 1 & 0 / 0        & 0 / 0       & 0 / 0     & 0 / 0      & 0 / 0      & 47,524 / 1 (0.00\%)      \\
            & Pick Order Item & 0 / 0 & 0 / 0    & 0 / 0      & 0 / 0        & 0 / 0        & 0 / 0   & 46,124 / 603,468 & 0 / 0       & 0 / 0     & 0 / 0      & 0 / 0      & 46,124 / 603,468 (1308.36\%) \\
            & Create Delivery Note  & 0 / 0 & 0 / 0      & 0 / 0        & 0 / 0          & 0 / 0          & 0 / 0     & 0 / 0          & 46,414 / 30,626 & 538 / 30,626 & 0 / 0        & 0 / 0        & 46,952 / 30,626 (65.23\%)  \\
            & Generate Group Delivery Note  & 0 / 0   & 0 / 0      & 0 / 0        & 0 / 0          & 0 / 0          & 0 / 0     & 0 / 0          & 0 / 0         & 0 / 0       & 0 / 0        & 24 / 46,945   & 24 / 46,945 (195,604.17\%)     \\
            & Deliver Order & 0 / 0   & 0 / 0      & 0 / 0        & 0 / 0          & 0 / 0          & 0 / 0     & 0 / 0          & 0 / 0         & 0 / 0       & 40,967 / 1,156 & 0 / 0        & 40,967 / 1,156 (2.28\%)   \\
            & Post Invoice & 0 / 0   & 0 / 0      & 0 / 0        & 0 / 0          & 0 / 0          & 0 / 0     & 0 / 0          & 0 / 0         & 0 / 0       & 0 / 0        & 63,758 / 1,798 & 63,758 / 1,798 (2.28\%)   \\
            & End     & 0 / 0   & 0 / 0      & 0 / 0        & 0 / 0          & 0 / 0          & 0 / 0     & 0 / 0          & 0 / 0         & 0 / 0       & 0 / 0        & 0 / 0        & 0 / 0 (0.00\%)          \\
        \bottomrule
    \end{tabular}
\end{adjustbox}

%% file: sections/40_evaluation/expected_unexpected_table_noreja.tex
\renewcommand{\arraystretch}{2.3}
\begin{adjustbox}{angle=90, center, caption={Ratio between the Expected Relationships Quantity and theUnexpected Relationships Quantity (Expected / Unexpected) of the \nth{3} \gls{aceg} Aggregation Level created by Noreja Approach.\label{tab:noreja_expected_unexpected_quantity}}, nofloat=table}
    \centering
    \tiny
    \begin{tabular}{p{0.05cm}p{1cm}p{0.6cm}p{0.8cm}p{0.9cm}p{1cm}p{1cm}p{0.8cm}p{1cm}p{1cm}p{0.8cm}p{0.8cm}p{0.8cm}p{1.4cm}}
        \toprule
            &         & \multicolumn{12}{c}{Target of Relationship}                                                                                                                           \\
            &         & Start & Receive Order   & Extract Order Item     & Prepare Order Delivery    & Plan Order Item Picking         & Pick Order Item   & Create Delivery Note        & Generate Group Delivery Note        & Deliver Order       & Post Invoice    & End       & Total           \\
        \midrule
        \multirow{17}{*}{\STAB{\rotatebox[origin=c]{90}{Source of Relationship}}}
            & Start   & 0 / 0 & 69,999 / 1 & 0 / 0       & 0 / 0      & 0 / 0           & 0 / 0     & 0 / 0         & 0 / 0         & 0 / 0         & 0 / 0      & 0 / 0     & 69,999 / 1 (0.00\%)       \\
            & Receive Order & 0 / 0 & 0 / 0     & 69,438 / 561 & 0 / 0      & 0 / 0           & 0 / 0     & 0 / 0         & 0 / 0         & 0 / 0         & 0 / 0      & 0 / 0     & 69,438 / 561 (0.81\%)    \\
            & Extract Order Item & 0 / 0 & 0 / 0     & 0 / 0       & 472,147 / 0 & 0 / 0           & 0 / 0     & 0 / 0         & 0 / 0         & 0 / 0         & 0 / 0      & 0 / 0     & 472,147 / 0 (0.00\%)      \\
            & Prepare Order Delivery & 0 / 0 & 0 / 0     & 0 / 0       & 0 / 0      & 356,779 / 116,184 & 0 / 0     & 0 / 0         & 0 / 0         & 0 / 0         & 0 / 0      & 0 / 0     & 356,779 / 116,184 (32.56\%) \\
            & Plan Order Item Picking & 0 / 0 & 0 / 0     & 0 / 0       & 0 / 0      & 0 / 0           & 47,525 / 0 & 0 / 0         & 0 / 0         & 0 / 0         & 0 / 0      & 0 / 0     & 47,525 / 0 (0.00\%)       \\
            & Pick Order Item & 0 / 0 & 0 / 0     & 0 / 0       & 0 / 0      & 0 / 0           & 0 / 0     & 646,963 / 2,629 & 0 / 0         & 0 / 0         & 0 / 0      & 0 / 0     & 646,963 / 2,629 (0.41\%)   \\
            & Create Delivery Note  & 0 / 0 & 0 / 0     & 0 / 0       & 0 / 0      & 0 / 0           & 0 / 0     & 0 / 0         & 46,971 / 35,465 & 42,112 / 35,465 & 0 / 0      & 0 / 0     & 89,083 / 35,465 (39.81\%)   \\
            & Generate Group Delivery Note  & 0 / 0 & 0 / 0     & 0 / 0       & 0 / 0      & 0 / 0           & 0 / 0     & 0 / 0         & 0 / 0         & 0 / 0         & 0 / 0      & 46,969 / 0 & 46,969 / 0 (0.00\%)       \\
            & Deliver Order & 0 / 0 & 0 / 0     & 0 / 0       & 0 / 0      & 0 / 0           & 0 / 0     & 0 / 0         & 0 / 0         & 0 / 0         & 42,093 / 30 & 0 / 0     & 42,093 / 30 (0.07\%)      \\
            & Post Invoice & 0 / 0 & 0 / 0     & 0 / 0       & 0 / 0      & 0 / 0           & 0 / 0     & 0 / 0         & 0 / 0         & 0 / 0         & 0 / 0      & 65,556 / 0 & 65,556 / 0 (0.00\%)       \\
            & End     & 0 / 0 & 0 / 0     & 0 / 0       & 0 / 0      & 0 / 0           & 0 / 0     & 0 / 0         & 0 / 0         & 0 / 0         & 0 / 0      & 0 / 0     & 0 / 0 (0.00\%)           \\   
        \bottomrule                        
    \end{tabular}
\end{adjustbox}

%% file: sections/40_evaluation/noviolation_violation_table_noreja.tex
\renewcommand{\arraystretch}{2.3}
\begin{adjustbox}{angle=90, center, caption={Ration between Quantity without Temporal Violation and Quantity with Temporal Violation (Quantity without Violation / Quantity with Violation) of the Aggregation Level 3 \gls{aceg} created by Noreja Approach.\label{tab:quantity_withoutviolation_violations}}, nofloat=table}
    \centering
    \tiny
    \begin{tabular}{p{0.05cm}p{1cm}p{0.6cm}p{0.8cm}p{0.9cm}p{1cm}p{1cm}p{0.8cm}p{1cm}p{1cm}p{0.8cm}p{0.8cm}p{0.8cm}p{1.4cm}}
        \toprule
            &         & \multicolumn{12}{c}{Target of Relationship}                                                                                                                \\
            &         & Start & Receive Order   & Extract Order Item   & Prepare Order Delivery        & Plan Order Item Picking    & Pick Order Item   & Create Delivery Note       & Generate Group Delivery Note    & Deliver Order    & Post Invoice    & End        & Total (\%)         \\
        \midrule
        \multirow{17}{*}{\STAB{\rotatebox[origin=c]{90}{Source of Relationship}}}
            & Start   & 0 / 0 & 69,999 / 0 & 0 / 0     & 0 / 0          & 0 / 0      & 0 / 0     & 0 / 0        & 1 / 0       & 0 / 0      & 0 / 0      & 0 / 0      & 70,000 / 0 (0.00\%)     \\
            & Receive Order & 0 / 0 & 0 / 0     & 69,438 / 0 & 0 / 0          & 0 / 0      & 0 / 0     & 0 / 0        & 0 / 0     & 0 / 0      & 0 / 0      & 561 / 0    & 69,999 / 0 (0.00\%)      \\
            & Extract Order Item & 0 / 0 & 0 / 0     & 0 / 0     & 461,962 / 10,185 & 0 / 0      & 0 / 0     & 0 / 0        & 0 / 0     & 0 / 0      & 0 / 0      & 0 / 0      & 461,962 / 10,185 (2.20\%) \\
            & Prepare Order Delivery & 0 / 0 & 0 / 0     & 0 / 0     & 0 / 0          & 356,779 / 0 & 0 / 0     & 68,057 / 1,035 & 0 / 0     & 23 / 0     & 0 / 0      & 47,069  / 0 & 471,928 / 1,035 (0.22\%)  \\
            & Plan Order Item Picking & 0 / 0 & 0 / 0     & 0 / 0     & 0 / 0          & 0 / 0      & 44,486 / 0 & 0 / 0        & 0 / 0     & 0 / 0      & 0 / 0      & 0 / 0      & 44,486 / 0 (0.00\%)      \\
            & Pick Order Item & 0 / 0 & 0 / 0     & 0 / 0     & 0 / 0          & 0 / 0      & 0 / 0     & 646,963 / 0   & 0 / 0     & 0 / 0      & 0 / 0      & 2,629 / 0   & 649,592 / 0 (0.00\%)     \\
            & Create Delivery Note  & 0 / 0 & 0 / 0     & 0 / 0     & 0 / 0          & 0 / 0      & 0 / 0     & 0 / 0        & 46,970 / 0 & 42,112 / 13 & 35,448 / 0  & 17 / 0     & 124,548 / 13 (0.01\%)    \\
            & Generate Group Delivery Note  & 0 / 0 & 0 / 0     & 0 / 0     & 0 / 0          & 0 / 0      & 0 / 0     & 0 / 0        & 0 / 0     & 0 / 0      & 0 / 0      & 46,969 / 0  & 46,969 / 0 (0.00\%)      \\
            & Deliver Order & 0 / 0 & 0 / 0     & 0 / 0     & 0 / 0          & 0 / 0      & 0 / 0     & 0 / 0        & 0 / 0     & 0 / 0      & 42,034 / 69 & 30 / 0     & 42,064 / 30  (0.07\%)    \\
            & Post Invoice & 0 / 0 & 0 / 0     & 0 / 0     & 0 / 0          & 0 / 0      & 0 / 0     & 0 / 0        & 0 / 0     & 0 / 0      & 0 / 0      & 65,556 / 0  & 65,556 / 0 (0.00\%)      \\
            & End     & 0 / 0 & 0 / 0     & 0 / 0     & 0 / 0          & 0 / 0      & 0 / 0     & 0 / 0        & 0 / 0     & 0 / 0      & 0 / 0      & 0 / 0      & 0 / 0 (0.00\%)          \\
        \bottomrule
    \end{tabular}
\end{adjustbox}    

%% file: main.bbl
\begin{thebibliography}{43}
\expandafter\ifx\csname natexlab\endcsname\relax\def\natexlab#1{#1}\fi
\providecommand{\url}[1]{\texttt{#1}}
\providecommand{\href}[2]{#2}
\providecommand{\path}[1]{#1}
\providecommand{\DOIprefix}{doi:}
\providecommand{\ArXivprefix}{arXiv:}
\providecommand{\URLprefix}{URL: }
\providecommand{\Pubmedprefix}{pmid:}
\providecommand{\doi}[1]{\href{http://dx.doi.org/#1}{\path{#1}}}
\providecommand{\Pubmed}[1]{\href{pmid:#1}{\path{#1}}}
\providecommand{\bibinfo}[2]{#2}
\ifx\xfnm\relax \def\xfnm[#1]{\unskip,\space#1}\fi
\bibitem[{van~der Aalst(2016)}]{DBLP:books/sp/Aalst16}
\bibinfo{author}{W.~M.~P. van~der Aalst}, \bibinfo{title}{Process Mining - Data
  Science in Action, Second Edition}, \bibinfo{publisher}{Springer},
  \bibinfo{year}{2016}.
\bibitem[{Leemans et~al.(2018)Leemans, Fahland, and van~der
  Aalst}]{leemans2018scalable}
\bibinfo{author}{S.~J. Leemans}, \bibinfo{author}{D.~Fahland},
  \bibinfo{author}{W.~M.~P. van~der Aalst},
\newblock \bibinfo{title}{Scalable process discovery and conformance checking},
\newblock \bibinfo{journal}{Software \& Systems Modeling} \bibinfo{volume}{17}
  (\bibinfo{year}{2018}) \bibinfo{pages}{599--631}.
\bibitem[{Buijs et~al.(2014)Buijs, van Dongen, and van~der
  Aals}]{buijs2014quality}
\bibinfo{author}{J.~C. Buijs}, \bibinfo{author}{B.~F. van Dongen},
  \bibinfo{author}{W.~M.~P. van~der Aals},
\newblock \bibinfo{title}{Quality dimensions in process discovery: The
  importance of fitness, precision, generalization and simplicity},
\newblock \bibinfo{journal}{International Journal of Cooperative Information
  Systems} \bibinfo{volume}{23} (\bibinfo{year}{2014})
  \bibinfo{pages}{1440001}.
\bibitem[{Augusto et~al.(2019)Augusto, Conforti, Dumas, La~Rosa, and
  Polyvyanyy}]{augusto2019split}
\bibinfo{author}{A.~Augusto}, \bibinfo{author}{R.~Conforti},
  \bibinfo{author}{M.~Dumas}, \bibinfo{author}{M.~La~Rosa},
  \bibinfo{author}{A.~Polyvyanyy},
\newblock \bibinfo{title}{Split miner: automated discovery of accurate and
  simple business process models from event logs},
\newblock \bibinfo{journal}{Knowledge and Information Systems}
  \bibinfo{volume}{59} (\bibinfo{year}{2019}) \bibinfo{pages}{251--284}.
\bibitem[{Augusto et~al.(2018)Augusto, Conforti, Dumas, La~Rosa, Maggi,
  Marrella, Mecella, and Soo}]{augusto2018automated}
\bibinfo{author}{A.~Augusto}, \bibinfo{author}{R.~Conforti},
  \bibinfo{author}{M.~Dumas}, \bibinfo{author}{M.~La~Rosa},
  \bibinfo{author}{F.~M. Maggi}, \bibinfo{author}{A.~Marrella},
  \bibinfo{author}{M.~Mecella}, \bibinfo{author}{A.~Soo},
\newblock \bibinfo{title}{Automated discovery of process models from event
  logs: Review and benchmark},
\newblock \bibinfo{journal}{IEEE Transactions on Knowledge and Data
  Engineering} \bibinfo{volume}{31} (\bibinfo{year}{2018})
  \bibinfo{pages}{686--705}.
\bibitem[{van~der Aalst(2011)}]{Aalst2011OnTR}
\bibinfo{author}{W.~M.~P. van~der Aalst},
\newblock \bibinfo{title}{On the representational bias in process mining},
\newblock in: \bibinfo{editor}{S.~Reddy}, \bibinfo{editor}{S.~Tata} (Eds.),
  \bibinfo{booktitle}{20th {IEEE} International Workshops on Enabling
  Technologies: Infrastructures for Collaborative Enterprises, {WETICE} 2011,
  Paris, France, 27-29 June 2011, Proceedings}, \bibinfo{publisher}{{IEEE}
  Computer Society}, \bibinfo{year}{2011}, pp. \bibinfo{pages}{2--7}.
  \URLprefix \url{https://doi.org/10.1109/WETICE.2011.64}.
  \DOIprefix\doi{10.1109/WETICE.2011.64}.
\bibitem[{van~der Aalst et~al.(2011)van~der Aalst, Adriansyah, De~Medeiros,
  Arcieri, Baier, Blickle, Bose, Van Den~Brand, Brandtjen, Buijs
  et~al.}]{van2011process}
\bibinfo{author}{W.~van~der Aalst}, \bibinfo{author}{A.~Adriansyah},
  \bibinfo{author}{A.~K.~A. De~Medeiros}, \bibinfo{author}{F.~Arcieri},
  \bibinfo{author}{T.~Baier}, \bibinfo{author}{T.~Blickle},
  \bibinfo{author}{J.~C. Bose}, \bibinfo{author}{P.~Van Den~Brand},
  \bibinfo{author}{R.~Brandtjen}, \bibinfo{author}{J.~Buijs}, et~al.,
\newblock \bibinfo{title}{Process mining manifesto},
\newblock in: \bibinfo{booktitle}{International Conference on Business Process
  Management}, \bibinfo{organization}{Springer}, \bibinfo{year}{2011}, pp.
  \bibinfo{pages}{169--194}.
\bibitem[{Gerke et~al.(2009)Gerke, Mendling, and Tarmyshov}]{gerke2009case}
\bibinfo{author}{K.~Gerke}, \bibinfo{author}{J.~Mendling},
  \bibinfo{author}{K.~Tarmyshov},
\newblock \bibinfo{title}{Case construction for mining supply chain processes},
\newblock in: \bibinfo{booktitle}{International Conference on Business
  Information Systems}, \bibinfo{organization}{Springer}, \bibinfo{year}{2009},
  pp. \bibinfo{pages}{181--192}.
\bibitem[{Lu et~al.(2015)Lu, Nagelkerke, van~de Wiel, and
  Fahland}]{DBLP:journals/tsc/LuNWF15}
\bibinfo{author}{X.~Lu}, \bibinfo{author}{M.~Nagelkerke},
  \bibinfo{author}{D.~van~de Wiel}, \bibinfo{author}{D.~Fahland},
\newblock \bibinfo{title}{Discovering interacting artifacts from {ERP}
  systems},
\newblock \bibinfo{journal}{{IEEE} Trans. Serv. Comput.} \bibinfo{volume}{8}
  (\bibinfo{year}{2015}) \bibinfo{pages}{861--873}.
\bibitem[{van~der Aalst(2019)}]{DBLP:conf/sefm/Aalst19}
\bibinfo{author}{W.~M.~P. van~der Aalst},
\newblock \bibinfo{title}{Object-centric process mining: Dealing with
  divergence and convergence in event data},
\newblock in: \bibinfo{editor}{P.~C. {\"{O}}lveczky},
  \bibinfo{editor}{G.~Sala{\"{u}}n} (Eds.), \bibinfo{booktitle}{Software
  Engineering and Formal Methods - 17th International Conference, {SEFM} 2019,
  Oslo, Norway, September 18-20, 2019, Proceedings}, volume
  \bibinfo{volume}{11724} of \textit{\bibinfo{series}{Lecture Notes in Computer
  Science}}, \bibinfo{publisher}{Springer}, \bibinfo{year}{2019}, pp.
  \bibinfo{pages}{3--25}. \URLprefix
  \url{https://doi.org/10.1007/978-3-030-30446-1\_1}.
  \DOIprefix\doi{10.1007/978-3-030-30446-1\_1}.
\bibitem[{Li et~al.(2018)Li, de~Murillas, de~Carvalho, and van~der
  Aalst}]{DBLP:conf/caise/LiMCA18}
\bibinfo{author}{G.~Li}, \bibinfo{author}{E.~G.~L. de~Murillas},
  \bibinfo{author}{R.~M. de~Carvalho}, \bibinfo{author}{W.~M.~P. van~der
  Aalst},
\newblock \bibinfo{title}{Extracting object-centric event logs to support
  process mining on databases},
\newblock in: \bibinfo{booktitle}{CAiSE Forum}, volume \bibinfo{volume}{317} of
  \textit{\bibinfo{series}{Lecture Notes in Business Information Processing}},
  \bibinfo{publisher}{Springer}, \bibinfo{year}{2018}, pp.
  \bibinfo{pages}{182--199}.
\bibitem[{de~Murillas et~al.(2020)de~Murillas, Reijers, and van~der
  Aalst}]{DBLP:journals/kais/MurillasRA20}
\bibinfo{author}{E.~G.~L. de~Murillas}, \bibinfo{author}{H.~A. Reijers},
  \bibinfo{author}{W.~M.~P. van~der Aalst},
\newblock \bibinfo{title}{Case notion discovery and recommendation: automated
  event log building on databases},
\newblock \bibinfo{journal}{Knowl. Inf. Syst.} \bibinfo{volume}{62}
  (\bibinfo{year}{2020}) \bibinfo{pages}{2539--2575}.
\bibitem[{Berti and van~der Aalst(2020)}]{10.1007/978-3-030-46633-6_2}
\bibinfo{author}{A.~Berti}, \bibinfo{author}{W.~van~der Aalst},
\newblock \bibinfo{title}{Extracting multiple viewpoint models from relational
  databases},
\newblock in: \bibinfo{editor}{P.~Ceravolo}, \bibinfo{editor}{M.~van Keulen},
  \bibinfo{editor}{M.~T. G{\'o}mez-L{\'o}pez} (Eds.),
  \bibinfo{booktitle}{Data-Driven Process Discovery and Analysis},
  \bibinfo{publisher}{Springer International Publishing},
  \bibinfo{address}{Cham}, \bibinfo{year}{2020}, pp. \bibinfo{pages}{24--51}.
\bibitem[{Andrews et~al.(2020)Andrews, {van Dun}, Wynn, Kratsch, Röglinger,
  and {ter Hofstede}}]{ANDREWS2020113265}
\bibinfo{author}{R.~Andrews}, \bibinfo{author}{C.~{van Dun}},
  \bibinfo{author}{M.~Wynn}, \bibinfo{author}{W.~Kratsch},
  \bibinfo{author}{M.~Röglinger}, \bibinfo{author}{A.~{ter Hofstede}},
\newblock \bibinfo{title}{Quality-informed semi-automated event log generation
  for process mining},
\newblock \bibinfo{journal}{Decision Support Systems} \bibinfo{volume}{132}
  (\bibinfo{year}{2020}) \bibinfo{pages}{113265}. \URLprefix
  \url{https://www.sciencedirect.com/science/article/pii/S0167923620300208}.
  \DOIprefix\doi{https://doi.org/10.1016/j.dss.2020.113265}.
\bibitem[{Esser and Fahland(2021)}]{DBLP:journals/corr/abs-2005-14552}
\bibinfo{author}{S.~Esser}, \bibinfo{author}{D.~Fahland},
\newblock \bibinfo{title}{Multi-dimensional event data in graph databases},
\newblock \bibinfo{journal}{J. Data Semant.} \bibinfo{volume}{10}
  (\bibinfo{year}{2021}) \bibinfo{pages}{109--141}. \URLprefix
  \url{https://doi.org/10.1007/s13740-021-00122-1}.
  \DOIprefix\doi{10.1007/s13740-021-00122-1}.
\bibitem[{Pearl(2019)}]{pearl2019seven}
\bibinfo{author}{J.~Pearl},
\newblock \bibinfo{title}{The seven tools of causal inference, with reflections
  on machine learning},
\newblock \bibinfo{journal}{Comm. of the ACM} \bibinfo{volume}{62}
  (\bibinfo{year}{2019}) \bibinfo{pages}{54--60}.
\bibitem[{van~der Aalst(2019)}]{van2019practitioner}
\bibinfo{author}{W.~M. van~der Aalst},
\newblock \bibinfo{title}{A practitioner’s guide to process mining:
  Limitations of the directly-follows graph},
\newblock \bibinfo{journal}{Procedia Computer Science} \bibinfo{volume}{164}
  (\bibinfo{year}{2019}) \bibinfo{pages}{321--328}.
\bibitem[{Jans et~al.(2019)Jans, Soffer, and Jouck}]{jans2019building}
\bibinfo{author}{M.~Jans}, \bibinfo{author}{P.~Soffer},
  \bibinfo{author}{T.~Jouck},
\newblock \bibinfo{title}{Building a valuable event log for process mining: an
  experimental exploration of a guided process},
\newblock \bibinfo{journal}{Enterprise Information Systems}
  \bibinfo{volume}{13} (\bibinfo{year}{2019}) \bibinfo{pages}{601--630}.
\bibitem[{Ciccio et~al.(2018)Ciccio, Maggi, Montali, and
  Mendling}]{DBLP:journals/is/CiccioMMM18}
\bibinfo{author}{C.~D. Ciccio}, \bibinfo{author}{F.~M. Maggi},
  \bibinfo{author}{M.~Montali}, \bibinfo{author}{J.~Mendling},
\newblock \bibinfo{title}{On the relevance of a business constraint to an event
  log},
\newblock \bibinfo{journal}{Inf. Syst.} \bibinfo{volume}{78}
  (\bibinfo{year}{2018}) \bibinfo{pages}{144--161}.
\bibitem[{Gerke et~al.(2009)Gerke, Claus, and Mendling}]{gerke2009process}
\bibinfo{author}{K.~Gerke}, \bibinfo{author}{A.~Claus},
  \bibinfo{author}{J.~Mendling},
\newblock \bibinfo{title}{Process mining of rfid-based supply chains},
\newblock in: \bibinfo{booktitle}{2009 IEEE Conference on Commerce and
  Enterprise Computing}, \bibinfo{organization}{IEEE}, \bibinfo{year}{2009},
  pp. \bibinfo{pages}{285--292}.
\bibitem[{Waibel et~al.(2020)Waibel, Novak, Bala, Revoredo, and
  Mendling}]{DBLP:conf/icpm/WaibelNBRM20}
\bibinfo{author}{P.~Waibel}, \bibinfo{author}{C.~Novak},
  \bibinfo{author}{S.~Bala}, \bibinfo{author}{K.~Revoredo},
  \bibinfo{author}{J.~Mendling},
\newblock \bibinfo{title}{Analysis of business process batching using causal
  event models},
\newblock in: \bibinfo{editor}{S.~J.~J. Leemans}, \bibinfo{editor}{H.~Leopold}
  (Eds.), \bibinfo{booktitle}{Process Mining Workshops - {ICPM} 2020
  International Workshops, Padua, Italy, October 5-8, 2020, Revised Selected
  Papers}, volume \bibinfo{volume}{406} of \textit{\bibinfo{series}{Lecture
  Notes in Business Information Processing}}, \bibinfo{publisher}{Springer},
  \bibinfo{year}{2020}, pp. \bibinfo{pages}{17--29}. \URLprefix
  \url{https://doi.org/10.1007/978-3-030-72693-5\_2}.
  \DOIprefix\doi{10.1007/978-3-030-72693-5\_2}.
\bibitem[{Weber et~al.(2015)Weber, Farshchi, Mendling, and
  Schneider}]{DBLP:conf/sac/WeberFMS15}
\bibinfo{author}{I.~Weber}, \bibinfo{author}{M.~Farshchi},
  \bibinfo{author}{J.~Mendling}, \bibinfo{author}{J.~Schneider},
\newblock \bibinfo{title}{Mining processes with multi-instantiation},
\newblock in: \bibinfo{booktitle}{Proceedings of the 30th Annual {ACM}
  Symposium on Applied Computing}, \bibinfo{publisher}{{ACM}},
  \bibinfo{year}{2015}, pp. \bibinfo{pages}{1231--1237}.
\bibitem[{van~der Aalst and Berti(2020)}]{DBLP:journals/fuin/AalstB20}
\bibinfo{author}{W.~M.~P. van~der Aalst}, \bibinfo{author}{A.~Berti},
\newblock \bibinfo{title}{Discovering object-centric petri nets},
\newblock \bibinfo{journal}{Fundam. Informaticae} \bibinfo{volume}{175}
  (\bibinfo{year}{2020}) \bibinfo{pages}{1--40}. \URLprefix
  \url{https://doi.org/10.3233/FI-2020-1946}.
  \DOIprefix\doi{10.3233/FI-2020-1946}.
\bibitem[{Li et~al.(2018)Li, Medeiros~de Carvalho, and van~der Aalst}]{8456350}
\bibinfo{author}{G.~Li}, \bibinfo{author}{R.~Medeiros~de Carvalho},
  \bibinfo{author}{W.~M. van~der Aalst},
\newblock \bibinfo{title}{Configurable event correlation for process discovery
  from object-centric event data},
\newblock in: \bibinfo{booktitle}{2018 IEEE International Conference on Web
  Services (ICWS)}, \bibinfo{year}{2018}, pp. \bibinfo{pages}{203--210}.
  \DOIprefix\doi{10.1109/ICWS.2018.00033}.
\bibitem[{Fahland(2019)}]{DBLP:conf/apn/Fahland19}
\bibinfo{author}{D.~Fahland},
\newblock \bibinfo{title}{Describing behavior of processes with many-to-many
  interactions},
\newblock in: \bibinfo{booktitle}{Application and Theory of Petri Nets and
  Concurrency}, volume \bibinfo{volume}{11522} of
  \textit{\bibinfo{series}{LNCS}}, \bibinfo{publisher}{Springer},
  \bibinfo{year}{2019}, pp. \bibinfo{pages}{3--24}.
\bibitem[{Esser and Fahland(2019)}]{DBLP:conf/bpm/EsserF19}
\bibinfo{author}{S.~Esser}, \bibinfo{author}{D.~Fahland},
\newblock \bibinfo{title}{Storing and querying multi-dimensional process event
  logs using graph databases},
\newblock in: \bibinfo{booktitle}{BPM Workshops}, volume \bibinfo{volume}{362}
  of \textit{\bibinfo{series}{LNBIP}}, \bibinfo{publisher}{Springer},
  \bibinfo{year}{2019}, pp. \bibinfo{pages}{632--644}.
\bibitem[{Fahland(2022)}]{fahland2022process}
\bibinfo{author}{D.~Fahland},
\newblock \bibinfo{title}{Process mining over multiple behavioral dimensions
  with event knowledge graphs},
\newblock in: \bibinfo{booktitle}{Process Mining Handbook},
  \bibinfo{publisher}{Springer}, \bibinfo{year}{2022}, pp.
  \bibinfo{pages}{274--319}.
\bibitem[{Ghahfarokhi et~al.(2021)Ghahfarokhi, Berti, and van~der
  Aalst}]{DBLP:journals/corr/abs-2103-07184}
\bibinfo{author}{A.~F. Ghahfarokhi}, \bibinfo{author}{A.~Berti},
  \bibinfo{author}{W.~M.~P. van~der Aalst},
\newblock \bibinfo{title}{Process comparison using object-centric process
  cubes},
\newblock \bibinfo{journal}{CoRR} \bibinfo{volume}{abs/2103.07184}
  (\bibinfo{year}{2021}). \URLprefix \url{https://arxiv.org/abs/2103.07184}.
  \href{http://arxiv.org/abs/2103.07184}{{\tt arXiv:2103.07184}}.
\bibitem[{{Song} et~al.(2016){Song}, {Jacobsen}, {Ye}, and {Ma}}]{7095577}
\bibinfo{author}{W.~{Song}}, \bibinfo{author}{H.~{Jacobsen}},
  \bibinfo{author}{C.~{Ye}}, \bibinfo{author}{X.~{Ma}},
\newblock \bibinfo{title}{Process discovery from dependence-complete event
  logs},
\newblock \bibinfo{journal}{IEEE Transactions on Services Computing}
  \bibinfo{volume}{9} (\bibinfo{year}{2016}) \bibinfo{pages}{714--727}.
  \DOIprefix\doi{10.1109/TSC.2015.2426181}.
\bibitem[{Diamantini et~al.(2016)Diamantini, Genga, Potena, and van~der
  Aalst}]{DBLP:journals/eswa/DiamantiniGPA16}
\bibinfo{author}{C.~Diamantini}, \bibinfo{author}{L.~Genga},
  \bibinfo{author}{D.~Potena}, \bibinfo{author}{W.~M.~P. van~der Aalst},
\newblock \bibinfo{title}{Building instance graphs for highly variable
  processes},
\newblock \bibinfo{journal}{Expert Syst. Appl.} \bibinfo{volume}{59}
  (\bibinfo{year}{2016}) \bibinfo{pages}{101--118}.
\bibitem[{Lu et~al.(2017)Lu, Fahland, Andrews, Suriadi, Wynn, ter Hofstede, and
  van~der Aalst}]{DBLP:conf/otm/LuFASWHA17}
\bibinfo{author}{X.~Lu}, \bibinfo{author}{D.~Fahland},
  \bibinfo{author}{R.~Andrews}, \bibinfo{author}{S.~Suriadi},
  \bibinfo{author}{M.~T. Wynn}, \bibinfo{author}{A.~H.~M. ter Hofstede},
  \bibinfo{author}{W.~M.~P. van~der Aalst},
\newblock \bibinfo{title}{Semi-supervised log pattern detection and exploration
  using event concurrence and contextual information},
\newblock in: \bibinfo{booktitle}{{OTM} Conferences {(1)}}, volume
  \bibinfo{volume}{10573} of \textit{\bibinfo{series}{Lecture Notes in Computer
  Science}}, \bibinfo{publisher}{Springer}, \bibinfo{year}{2017}, pp.
  \bibinfo{pages}{154--174}.
\bibitem[{Leemans et~al.(2022)Leemans, van Zelst, and Lu}]{leemans2022partial}
\bibinfo{author}{S.~J. Leemans}, \bibinfo{author}{S.~J. van Zelst},
  \bibinfo{author}{X.~Lu},
\newblock \bibinfo{title}{Partial-order-based process mining: a survey and
  outlook},
\newblock \bibinfo{journal}{Knowledge and Information Systems}
  (\bibinfo{year}{2022}) \bibinfo{pages}{1--29}.
\bibitem[{Dumas and
  Garc{\'{\i}}a{-}Ba{\~{n}}uelos(2015)}]{DBLP:conf/apn/DumasG15}
\bibinfo{author}{M.~Dumas},
  \bibinfo{author}{L.~Garc{\'{\i}}a{-}Ba{\~{n}}uelos},
\newblock \bibinfo{title}{Process mining reloaded: Event structures as a
  unified representation of process models and event logs},
\newblock in: \bibinfo{booktitle}{Petri Nets}, volume \bibinfo{volume}{9115} of
  \textit{\bibinfo{series}{Lecture Notes in Computer Science}},
  \bibinfo{publisher}{Springer}, \bibinfo{year}{2015}, pp.
  \bibinfo{pages}{33--48}.
\bibitem[{Ponce-de Le{\'o}n et~al.(2015)Ponce-de Le{\'o}n, Rodr{\'i}guez,
  Carmona, Heljanko, and Haar}]{10.1007/978-3-319-24953-7_4}
\bibinfo{author}{H.~Ponce-de Le{\'o}n}, \bibinfo{author}{C.~Rodr{\'i}guez},
  \bibinfo{author}{J.~Carmona}, \bibinfo{author}{K.~Heljanko},
  \bibinfo{author}{S.~Haar},
\newblock \bibinfo{title}{Unfolding-based process discovery},
\newblock in: \bibinfo{editor}{B.~Finkbeiner}, \bibinfo{editor}{G.~Pu},
  \bibinfo{editor}{L.~Zhang} (Eds.), \bibinfo{booktitle}{Automated Technology
  for Verification and Analysis}, \bibinfo{publisher}{Springer International
  Publishing}, \bibinfo{address}{Cham}, \bibinfo{year}{2015}, pp.
  \bibinfo{pages}{31--47}.
\bibitem[{Bergenthum(2019)}]{DBLP:conf/icpm/Bergenthum19}
\bibinfo{author}{R.~Bergenthum},
\newblock \bibinfo{title}{Prime miner - process discovery using prime event
  structures},
\newblock in: \bibinfo{booktitle}{International Conference on Process Mining,
  {ICPM} 2019, Aachen, Germany, June 24-26, 2019}, \bibinfo{publisher}{{IEEE}},
  \bibinfo{year}{2019}, pp. \bibinfo{pages}{41--48}.
\bibitem[{Conforti et~al.(2016)Conforti, Dumas, Garc{\'{\i}}a{-}Ba{\~{n}}uelos,
  and Rosa}]{DBLP:journals/is/ConfortiDGR16}
\bibinfo{author}{R.~Conforti}, \bibinfo{author}{M.~Dumas},
  \bibinfo{author}{L.~Garc{\'{\i}}a{-}Ba{\~{n}}uelos}, \bibinfo{author}{M.~L.
  Rosa},
\newblock \bibinfo{title}{{BPMN} miner: Automated discovery of {BPMN} process
  models with hierarchical structure},
\newblock \bibinfo{journal}{Inf. Syst.} \bibinfo{volume}{56}
  (\bibinfo{year}{2016}) \bibinfo{pages}{284--303}.
\bibitem[{Calvanese et~al.(2019)Calvanese, Ghilardi, Gianola, Montali, and
  Rivkin}]{calvanese2019formal}
\bibinfo{author}{D.~Calvanese}, \bibinfo{author}{S.~Ghilardi},
  \bibinfo{author}{A.~Gianola}, \bibinfo{author}{M.~Montali},
  \bibinfo{author}{A.~Rivkin},
\newblock \bibinfo{title}{Formal modeling and smt-based parameterized
  verification of data-aware bpmn},
\newblock in: \bibinfo{booktitle}{International Conference on Business Process
  Management}, \bibinfo{organization}{Springer}, \bibinfo{year}{2019}, pp.
  \bibinfo{pages}{157--175}.
\bibitem[{Martin et~al.(2020)Martin, Depaire, Caris, and
  Schepers}]{martin2020retrieving}
\bibinfo{author}{N.~Martin}, \bibinfo{author}{B.~Depaire},
  \bibinfo{author}{A.~Caris}, \bibinfo{author}{D.~Schepers},
\newblock \bibinfo{title}{Retrieving the resource availability calendars of a
  process from an event log},
\newblock \bibinfo{journal}{Information Systems} \bibinfo{volume}{88}
  (\bibinfo{year}{2020}) \bibinfo{pages}{101463}.
\bibitem[{Augusto et~al.(2020)Augusto, Dumas, and
  Rosa}]{DBLP:conf/icpm/AugustoDR20}
\bibinfo{author}{A.~Augusto}, \bibinfo{author}{M.~Dumas},
  \bibinfo{author}{M.~L. Rosa},
\newblock \bibinfo{title}{Automated discovery of process models with true
  concurrency and inclusive choices},
\newblock in: \bibinfo{booktitle}{{ICPM} Workshops}, volume
  \bibinfo{volume}{406} of \textit{\bibinfo{series}{Lecture Notes in Business
  Information Processing}}, \bibinfo{publisher}{Springer},
  \bibinfo{year}{2020}, pp. \bibinfo{pages}{43--56}.
\bibitem[{Berti et~al.(2021)Berti, Li, Schuster, and van Zelst}]{pmtk}
\bibinfo{author}{A.~Berti}, \bibinfo{author}{C.-Y. Li},
  \bibinfo{author}{D.~Schuster}, \bibinfo{author}{S.~J. van Zelst},
\newblock \bibinfo{title}{The process mining toolkit (pmtk): Enabling advanced
  process mining in an integrated fashion (extended abstract)},
\newblock in: \bibinfo{editor}{G.~Kalenkova~A., Janssenswillen} (Ed.),
  \bibinfo{booktitle}{Proceedings of the ICPM Demo Track 2021, co-located with
  1st International Conference on Process Mining (ICPM 2021)},
  \bibinfo{year}{2021}.
\bibitem[{Berti et~al.(2019)Berti, van Zelst, and van~der
  Aalst}]{DBLP:journals/corr/abs-1905-06169}
\bibinfo{author}{A.~Berti}, \bibinfo{author}{S.~J. van Zelst},
  \bibinfo{author}{W.~M.~P. van~der Aalst},
\newblock \bibinfo{title}{Process mining for python (pm4py): Bridging the gap
  between process- and data science},
\newblock \bibinfo{journal}{CoRR} \bibinfo{volume}{abs/1905.06169}
  (\bibinfo{year}{2019}). \URLprefix \url{http://arxiv.org/abs/1905.06169}.
  \href{http://arxiv.org/abs/1905.06169}{{\tt arXiv:1905.06169}}.
\bibitem[{Grover and Leskovec(2016)}]{DBLP:conf/kdd/GroverL16}
\bibinfo{author}{A.~Grover}, \bibinfo{author}{J.~Leskovec},
\newblock \bibinfo{title}{node2vec: Scalable feature learning for networks},
\newblock in: \bibinfo{editor}{B.~Krishnapuram}, \bibinfo{editor}{M.~Shah},
  \bibinfo{editor}{A.~J. Smola}, \bibinfo{editor}{C.~C. Aggarwal},
  \bibinfo{editor}{D.~Shen}, \bibinfo{editor}{R.~Rastogi} (Eds.),
  \bibinfo{booktitle}{Proceedings of the 22nd {ACM} {SIGKDD} International
  Conference on Knowledge Discovery and Data Mining, San Francisco, CA, USA,
  August 13-17, 2016}, \bibinfo{publisher}{{ACM}}, \bibinfo{year}{2016}, pp.
  \bibinfo{pages}{855--864}. \URLprefix
  \url{https://doi.org/10.1145/2939672.2939754}.
  \DOIprefix\doi{10.1145/2939672.2939754}.
\bibitem[{Narayanan et~al.(2017)Narayanan, Chandramohan, Venkatesan, Chen, Liu,
  and Jaiswal}]{DBLP:journals/corr/NarayananCVCLJ17}
\bibinfo{author}{A.~Narayanan}, \bibinfo{author}{M.~Chandramohan},
  \bibinfo{author}{R.~Venkatesan}, \bibinfo{author}{L.~Chen},
  \bibinfo{author}{Y.~Liu}, \bibinfo{author}{S.~Jaiswal},
\newblock \bibinfo{title}{graph2vec: Learning distributed representations of
  graphs},
\newblock \bibinfo{journal}{CoRR} \bibinfo{volume}{abs/1707.05005}
  (\bibinfo{year}{2017}). \URLprefix \url{http://arxiv.org/abs/1707.05005}.
  \href{http://arxiv.org/abs/1707.05005}{{\tt arXiv:1707.05005}}.

\end{thebibliography}
